\documentclass[fleqn,usenatbib]{mnras}

\usepackage{newtxtext,newtxmath}

\usepackage[T1]{fontenc}
\usepackage{ae,aecompl}

\usepackage{graphicx}	
\usepackage{amsmath}	
\usepackage{pdflscape}

\newcommand{\halpha}{$\rm{H\rm{\alpha}}$}
\newcommand{\hbeta}{$\rm{H\rm{\beta}}$}
\newcommand{\oall}{$\text{[OIII]}\rm{\lambda}\rm{\lambda}4959,5007$}
\newcommand{\oiii}{$\text{OIII}$}
\newcommand{\ofive}{$\text{[OIII]}\rm{\lambda}5007$}
\newcommand{\ofour}{$\text{[OIII]}\rm{\lambda}4959$}
\newcommand{\ang}{\mbox{\normalfont\AA}}
\newcommand{\nii}{$\text{[NII]}\rm{\lambda}\rm{\lambda}6548,6563$}
\newcommand{\msun}{$\text{M}_{\odot}$}
\newcommand{\msig}{$\rm{M_{BH} - \sigma_{*}}$}
\newcommand{\lya}{$\rm{Ly\alpha}$}
\newcommand{\civ}{C\textsc{iv}}

\title[Black hole masses of RWGs]{The black hole masses of extremely luminous radio-WISE selected galaxies}

\author[E. R. Ferris et al.]{
E. R. Ferris,$^{1}$\thanks{E-mail: erf5@leicester.ac.uk}
A. W. Blain,$^{1}$
R.J. Assef,$^{2}$
N. A. Hatch,$^{3}$ 
A. Kimball,$^{4}$
M. Kim,$^{5}$ 
\newauthor
A. Sajina,$^{6}$
A. Silva,$^{7}$
D. Stern,$^{8}$
T. Diaz-Santos,$^{2}$
C-W. Tsai,$^{9}$
D. Wylezalek$^{10}$
\\
$^{1}$School of Physics and Astronomy, University of Leicester, University Road, LE1 7RH, UK\\
$^{2}$N\'ucleo de Astronom\'ia de la Facultad de Ingenier\'ia y Ciencias, Universidad Diego Portales, Av. Ej\'ercito Libertador 441, Santiago,\\ Chile, 8370109\\
$^{3}$School of Physics and Astronomy, University of Nottingham, University Park, Nottingham, NG7 2RD, UK\\
$^{4}$National Radio Astronomy Observatory, 1003 Lopezville Road, Socorro, NM 87801, USA\\
$^{5}$Department of Astronomy and Atmospheric Sciences, Kyungpook National University, Daegu 702-701, Korea\\
$^{6}$Department of Physics and Astronomy, Tufts University, 574 Boston Ave, Medford, MA 02155, USA\\
$^{7}$National astronomical observatory of Japan, National institutes of natural sciences, 2-21-1, Osawa, Mitaka, Tokyo 181-8588, Japan\\
$^{8}$Jet Propulsion Laboratory, California Institute of Technology, 4800 Oak Grove Drive, Pasadena, CA 91109, USA\\
$^{9}$National Astronomical Observatories, Chinese Academy of Sciences, 20A Datun Road, Chaoyang District, Beijing 100012, China\\
$^{10}$Astronomisches Rechen-Institut, Zentrum für Astronomie der Universität Heidelberg, Mönchhofstraße 12-14, 69120 Heidelberg, Germany}
\date{Accepted XXX. Received YYY; in original form ZZZ}

\pubyear{2019}

\begin{document}
\label{firstpage}
\pagerange{\pageref{firstpage}--\pageref{lastpage}}
\maketitle

\begin{abstract}
We present near-IR photometry and spectroscopy of 30 extremely luminous radio and mid-IR selected galaxies. With bolometric luminosities exceeding $\sim10^{13}$\,$\rm{L_{\odot}}$ and redshifts ranging from $z = 0.880-2.853$, we use VLT instruments X-shooter and ISAAC to investigate this unique population of galaxies. Broad multi-component emission lines are detected in 18 galaxies and we measure the near-IR lines \hbeta{}, \oall{} and \halpha{} in six, 15 and 13 galaxies respectively, with 10 \lya{} and five \civ{} lines additionally detected in the UVB arm. We use the broad \ofive{} emission lines as a proxy for the bolometric AGN luminosity, and derive lower limits to supermassive black hole masses of $10^{7.9}$-$10^{9.4}$\,\msun{} with expectations of corresponding host masses of $10^{10.4}$-$10^{12.0}$\,\msun{}.We measure $\rm{\lambda}_{Edd}$ > 1 for eight of these sources at a $2\sigma$ significance. Near-IR photometry and SED fitting are used to compare stellar masses directly. We detect both Balmer lines in five galaxies and use these to infer a mean visual extinction of $A_{V}$ = 2.68 mag. Due to non-detections and uncertainties in our \hbeta{} emission line measurements, we simulate a broad \hbeta{} line of FWHM = 1480\,$\rm{kms^{-1}}$ to estimate extinction for all sources with measured \halpha{} emission. We then use this to infer a mean $A_{V}=3.62$ mag, demonstrating the highly-obscured nature of these galaxies, with the consequence of increasing our estimates of black-hole masses by an 0.5 orders of magnitude in the most extreme and obscured cases. 
\end{abstract}




\section{Introduction}
The NASA Wide-field Infrared Survey Explorer (WISE) \citep{Wright2010} discovered $\sim$ 1000 rare, extremely luminous Infra-Red (IR) galaxies, selected using mid-IR colours, including Hot, Dust-Obscured Galaxies (Hot DOGs) \citep{Wu2012,Eisenhardt2012,Wu2014,Tsai2018,diazsantos18} and radio-selected WISE galaxies (RWGs) \citep{Lonsdale2015,Silva2015,Jones2015} which are comparably luminous but with more intense radio emission. These RWG Active Galactic Nuclei (AGNs) have been selected by combining mid-IR WISE and NRAO VLA Sky Survey (NVSS) radio data, with target selection detailed in \citet{Lonsdale2015}; also see section \ref{sec:select}. RWGs have compact radio emission still confined within their hosts, with young radio jets potentially exciting the Interstellar Medium (ISM). 

With Spectral Energy Distributions (SEDs) dominated by thermal emission from hot/warm dust in the optical to mid-IR, RWGs are likely radiatively efficient, in their peak fuelling phase and perhaps undergoing powerful feedback processes \citep{Hopkins2006}. They are highly-obscured systems that remain distinct from other very luminous active galaxy populations and provide a new opportunity to uncover some of the most extreme AGN in the universe, in their most dramatic black hole (BH) growth and feedback phase. RWGs' internal structure and morphologies may provide insight into a brief, key, peak phase in galaxy and quasar evolution. 

Follow up campaigns with the Atacama Large Millimeter/Submillimeter Array (ALMA) \citep{Lonsdale2015,Silva2015,Tsai2015,Diaz-santos2016,diazsantos18} have revealed a wide range of AGN spectral diagnostics, and complex morphologies in the ISM of both RWGs and similarly selected Hot DOGs. With well established radio galaxies apparently found in proto-clusters \citep{Wylezalek2013,Hatch14}, and the environments of RWGs showing a 4-6 times overdensity of ultraluminous, dusty, high-redshift 850\,-$\rm{\mu}$m -selected galaxies on the sky \citep[SMGs, e.g.][]{Blain2002,Casey2014} as compared with typical blank fields \citep{Jones2015,Silva2015}, and a more modest overdensity of red Spitzer-IRAC-selected galaxies \citep{Jordan}. This is potentially consistent with RWGs, and these ultra-luminous, dusty AGNs in general, being found in an active, extended still-virializing filamentary proto-cluster environments over scales in excess of five arcmin (3\,Mpc) consistent with theoretical models \citep[e.g.][]{Chiang17}. 

Our sources, being potential sites of the most powerful and intense AGN fuelling and feedback, provide a unique opportunity to study the AGN phenomenon in extremes, and to determine whether RWGs are distinct from other AGN classifications. This is the case for other WISE discovered sources including Extremely Luminous IR Galaxies (ELIRGs) \citep{Tsai2015,Tsai2018}. In addition, the extremely luminous nature of our sources ($\rm{L_{bol}} \sim 10^{47}$\,$\rm{erg\,s^{-1}}$), suggests that these luminous RWGs will be at their peak of activity. In a substantial increase in depth compared to previous large telescope IR spectroscopy of RWGs \citep{Kim2013}, we present high-resolution near-IR spectra of 27 RWGs, with near-IR imaging of 30 sources, taken in good seeing ($\sim 0\farcs8$), in order to determine the BH, host and galaxy stellar masses. Specifically, the availability of virial mass estimates, based on scaling relations matched to reverberation-mapped samples \citep{Kaspi2000}, and the relation between the stellar velocity dispersion of the host galaxy bulge and BH mass \citep[\msig,][]{Magorrian1998,Merrit2000,Gebhardt2000}, enables the calculation of the BH and host galaxy masses.

Defined as having masses $\geq10^{5}$\,\msun{}, supermassive BH masses can now be determined in various ways using high-resolution spectroscopy. Here we demonstrate two different approaches to determine RWG BH masses, using both a virial mass estimate based on the Balmer line properties and the tight correlation with the $L_{5100\,\ang{}}$ continuum luminosity \citep{Greene2005}, or an assumed Eddington ratio and the \ofive{} luminosity as a proxy for the bolometric luminosity. With the evolution of BHs and their hosts appearing closely linked, and AGN-driven outflows and feedback thought to be the crucial joining mechanism, these masses are vital properties to determine, and can test the validity of these scaling relations in the most dramatic cases.

The paper is set out as follows: Section \ref{sec:select} presents the selection criteria for our chosen targets. Section \ref{sec:obs} discusses the observation process and data reduction steps taken. Section \ref{sec:specprop} discusses the measured spectral properties of the emission lines and methods used. Section \ref{sec:bh} presents the main results of this work in terms of measured BH and host masses with discussions and conclusions presented in sections \ref{sec:discuss}, and \ref{sec:conc}, respectively.

The following cosmological parameters are used throughout this analysis: $H_{0} = 71\,\rm{kms^{-1}\,Mpc^{-1}}$, $\Omega_{\Lambda} = 0.73$ and $\Omega_{M} = 0.27$.

\subsection{Selection Criteria}
\label{sec:select}

The parent sample \citep{Lonsdale2015} was produced by cross-matching sources from the WISE All-sky catalogue \citep{Wright2010} with the NVSS radio survey \citep{Condon1998}, covering sources north of $\delta = -40^{\circ}$ to eliminate non-radio sources. A mid-IR colour criterion of $(W2 - W3) + 1.25\,(W1 - W2) > 7$ was then used to select the reddest sources in the WISE population, making use of the WISE colours, with an additional criterion of $\rm{-1 < \log  f_{22 \mu m}/f_{20 cm} < 0}$ added to avoid both radio-quiet and radio-loud galaxies: our sample is then radio intermediates. Rejecting sources within $\pm 10^{\circ}$ of the Galactic plane, the higher resolution of the Faint Images of the Radio Sky at Twenty centimetres (FIRST) radio survey \citep{Becker1995} was used for positional information, where available.
 \par
\begin{figure}
\centering
\includegraphics[width = 0.48\textwidth]{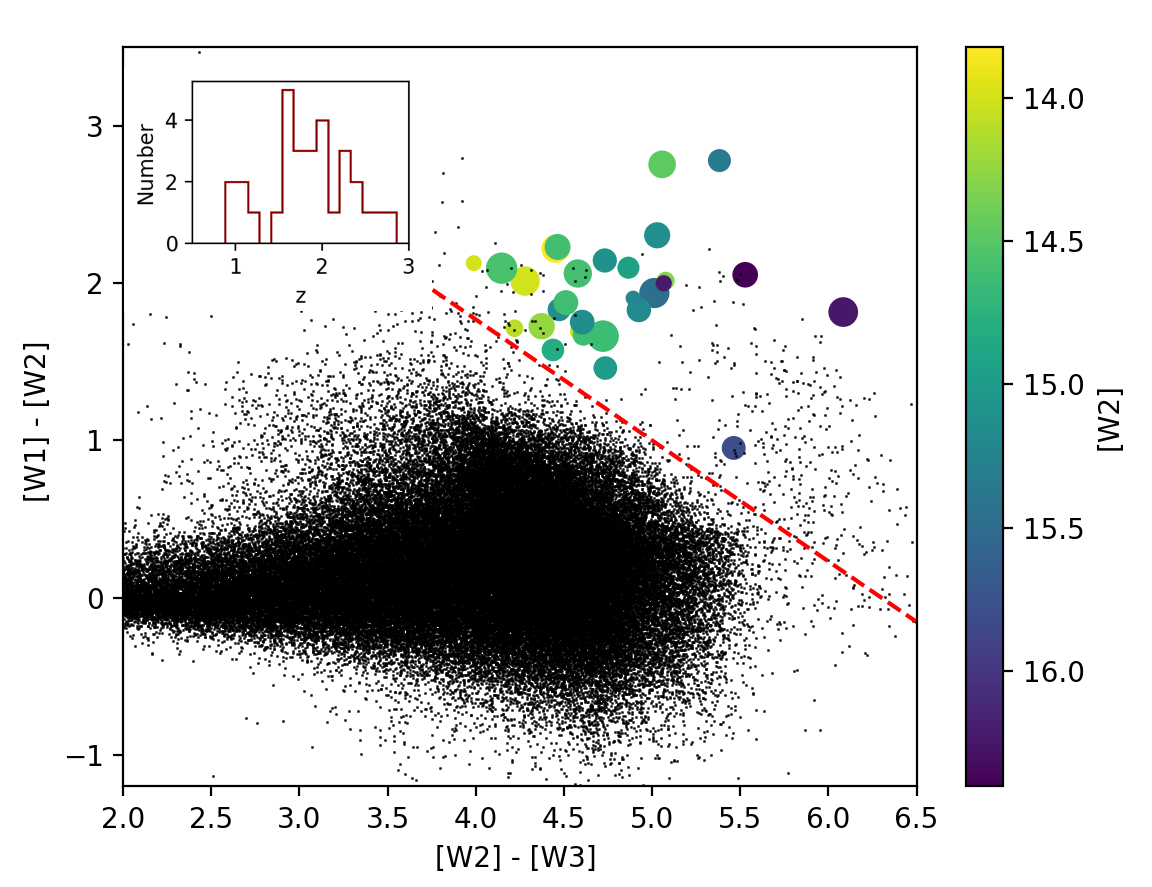}
\caption{\label{fig:select} WISE W1-W2-W3 colour-colour plot (in Vega magnitudes) showing the colours of the galaxies in our sample (crosses) compared to a random selection of 10,000 galaxies taken from the WISE All-sky data release. The dashed red line demonstrates the colour selection criterion: $(W2 - W3) + 1.25(W1 - W2) > 7$. Marker size is used to represent radio loudness with larger points being more radio loud. Marker shading represents W2 magnitude (Vega) as detailed in the colour bar on the right. The inserted histogram shows the redshift distribution of the 30 galaxies in our sample.}
\end{figure}

The location of our observed sample in WISE colour - colour space compared to 10,000 randomly selected sources taken from the WISE All-sky catalogue is shown in Fig. \ref{fig:select}. The dashed red line represents the mid-IR colour selection criteria \citep{Lonsdale2015} and demonstrates the ability to select all galaxies significantly redder than the main WISE population, and extremely bright in the W2 band. Rejection of extended radio-loud systems ensures we target young radio jets still confined to their hosts, confirmed by follow-up VLA snapshots \citep{lonsdale16}. 

Here, we targeted the reddest AGN in the mid/near-IR and therefore observe highly-obscured systems, likely within their peak fuelling phases \citep{lonsdale16}. Our VLT-derived redshift distribution of $z \approx 1-3$, as shown by the inserted histogram in Fig. \ref{fig:select}, means these AGNs span the peak epoch of BH and galaxy formation \citep{zakamska16}. 
\par
Observations of 30 available RWGs were made with the Very Large Telescope (VLT) Infrared Spectrometer and Array Camera (ISAAC) instrument with both the $J$ and $K_{s}$ bands in order to obtain stellar mass estimates via SED fitting. We obtained 27 spectra with X-shooter split across three arms, targeted using the $J$/$K_{s}$ band images. The overall sample was selected to consist of WISE W2-bright, highly-obscured, compact radio sources, that are the reddest radio-intermediate AGNs in the WISE population. All 30 targets have ALMA $\rm{870\,\mu m}$ observations from \citet{Lonsdale2015}, with flux densities detailed in Table \ref{tab:phot}, but were selected independent of that criterion.
 
\section{Observations and data reduction}
\label{sec:obs}
\subsection{Near-IR Photometry}
We obtained near-IR ($J$ and $K_{s}$ band) imaging using ISAAC \citep{Moowood1998} on VLT UT3 across nine observation nights in June 2013. Across the 30 objects, 29 were observed in the $K_{s}$ band, with 14 also observed in the $J$ band, the difference being due to time constraints. The observations were taken at low airmass ($\leq1.4$), with clear sky conditions and an average seeing of $\rm{0\farcs89}$. A total of $6\times15$\,s exposures were taken for each source in each band. Note that our observation period was close to the instrument being decommissioned, and one of the quadrants of the $1024\times1024$ pixel array was not working. This did not affect any of our observations as we centred on the apex of the L-shaped working region of the chip. Data reduction was done using the jitter recipe in the ISAAC data reduction pipeline.
\par
\begin{figure}
\centering
\includegraphics[width = 0.48\textwidth]{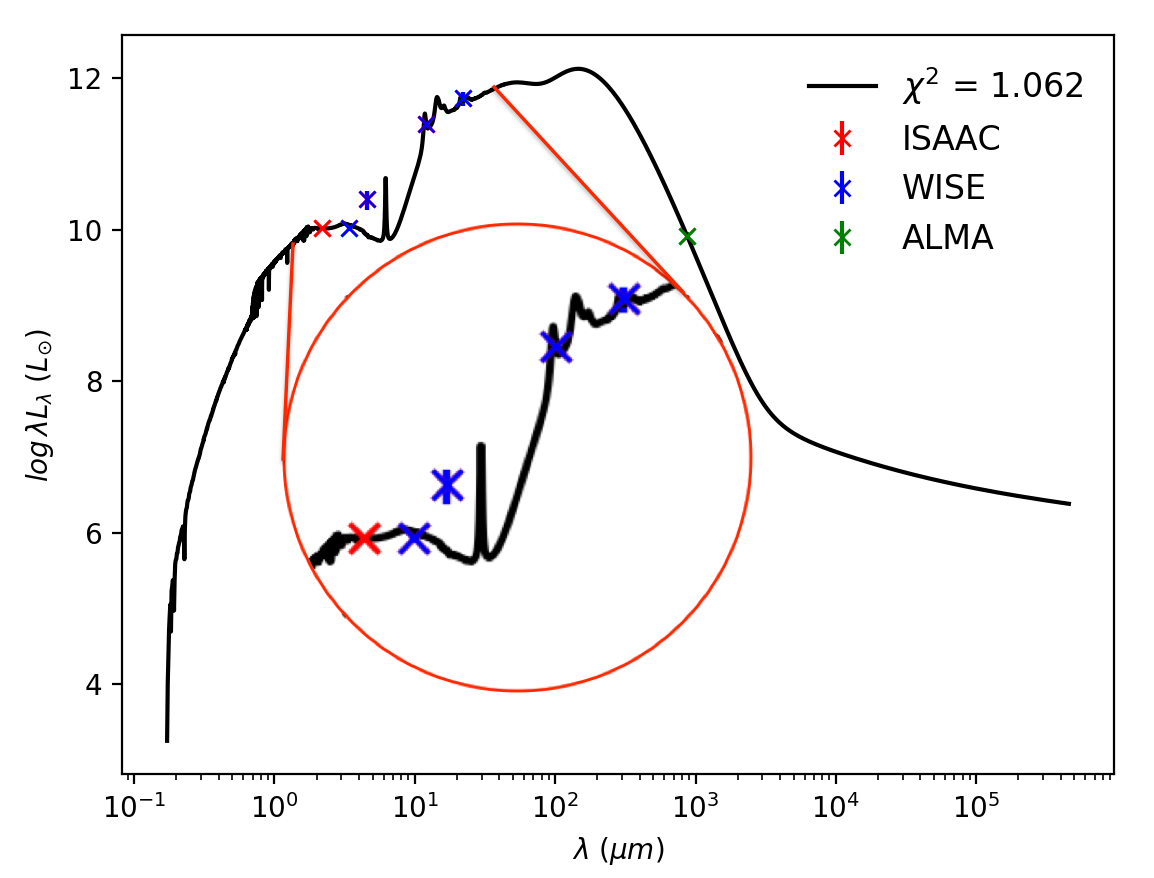}
\caption{\label{fig:sed} Example SED fitting of WISE J071433.54$-$363552, with a zoomed in red circle on our photometry points. This is the best fitting SED template for the ISAAC $K_{s}$, WISE W1, W2, W3, W4 and ALMA $\rm{870\,\mu m}$ photometry.}
\end{figure}

The SExtractor tool \citep{Bertin1996} was used to calculate flux densities in the $J$ and $K_{s}$ bands, with sources matched from the Two Micron All-Sky Survey (2MASS) catalogue \citep{2mass} used to determine median zero-point magnitudes for each band. Table \ref{tab:phot} provides the $J$ and $K_{s}$ band flux densities, and also those catalogued in the four WISE bands and ALMA $870\,\mu \rm{m}$ fluxes \citep{Lonsdale2015}. Stellar masses are estimated based on SED fitting in the seven wavebands, and provided with $\rm{\chi}^{2}_{\nu}$ values in Table \ref{tab:phot}. Note that available NVSS 1.4\,GHz flux densities are not included as they produced significantly worse SED fits due to the enhanced radio emission. The \texttt{MAGPHYS} SED-fitting code \citep{Dacunha2008} was used, providing best-fitting SEDs and their associated physical properties. Due to the lack of well-fitting IR SED templates for the galaxies in the RWG sample, \citep[see][]{Lonsdale2015}, $\rm{\chi}^{2}_{\nu}$ values are provided as a guide to the uncertainties in these estimates. In addition, for the sources that exhibit broad line emission and are therefore dominated by AGN emission, a lack of suitable components in the \texttt{MAGPHYS} SED templates adds increased uncertainty to our stellar mass estimates for these specific sources. These sources with detected broad lines are distinguished by their stellar masses being asterisked in Table \ref{tab:phot}. In these cases the stellar masses calculated are still provided but are likely overestimates of the true values. A typical example SED fit template is provided in Fig. \ref{fig:sed}, with a zoom in included on our photometric data points.

\subsection{Near-IR Spectroscopy}
In a consecutive three-night period straight after our ISAAC imaging, 27 sources were observed with X-shooter \citep{Vernet2011}, on VLT UT2. Simultaneous observations across the three wavelength arms; UVB, VIS, NIR, at slit widths $\rm{1\farcs6}$, $\rm{1\farcs5}$ and $\rm{1\farcs2}$ respectively, covered a full wavelength range of 300 - 2480\,nm. Sources with clear ISAAC imaging were selected to assist in correct slit placement and then confirmed by test guider observations. The targets were observed in nodding mode with a maximum distance of $\rm{5''}$ between nodding positions. A total integration time of 2$\times$600\,s was used for the NIR arm, 2$\times$400\,s for UVB and 2$\times$520\,s for VIS, with repeated ABBA nod sequences. The observations were done at low airmass ($\leq 1.16$) and with excellent seeing ranging from $0\farcs67 - 1\farcs15$. Following each science observation a standard star (LTT3218 or Feige110) was observed for telluric and relative flux calibration using standard response tables.
\par
The X-shooter Reflex pipeline\footnote{www.eso.org/sci/software/esoreflex} was used for data reduction. Flat-field corrections were not needed as observations were taken in nodding mode, and bias/dark frames were only required for UVB and VIS observations; the subtraction was done automatically for the NIR. Wavelength calibration was performed with an Th-Ar arc lamp; an average of 1716 lines were used to find a wavelength solution, with a mean error of $\sigma_{\rm RMS} = 0.09 \rm{\mathring A}$. Finally the IRAF \citep{IRAF} \texttt{dopcor} package was used to convert the spectra from observed to rest wavelength.
\par
The final science products include 2D and 1D spectra for each of the three arms, spanning a total wavelength range of 300 - 2480\,nm. The 2D spectra span $\rm{15\farcs75}$ in the spatial direction.

\begin{landscape}
\begin{table}
\centering
\caption{\label{tab:phot} The VLT determined redshift of each source along with ISAAC $J$ and $K_{s}$ band flux densities. Stellar masses are determined by SED fitting (see text for more details), based on the calculated ISAAC band flux densities along with those in the W1 (3.4 $\rm{\mu}$m), W2 (4.6 $\rm{\mu}$m), W3 (12 $\rm{\mu}$m) and W4 (22 $\rm{\mu}$m) WISE bands and ALMA (870 $\rm{\mu}$m) flux densities \citep{Lonsdale2015}. $\rm{\chi}^{2}_{\nu}$ values from the SED fitting are provided as a guide to the uncertainties in the stellar masses. Note: Asterisked source names were not observed by X-shooter. Asterisked redshifts of 2.00 are assumed where no spectroscopic redshift was available. Asterisked stellar mass values are likely overestimates of the true masses due these sources exhibiting broad line emission which is not suitably covered in the SED templates used.} 

\begin{tabular}{lllllllllll}
\hline
\hline
WISE designation& $z$&  $f_{\rm{J}}$ & $f_{\rm{Ks}}$ &$f_{\rm{W1}}$ & $f_{\rm{W2}}$ & $f_{\rm{W3}}$ & $f_{\rm{W4}}$ & $f_{\rm{ALMA\, 870\,\mu m}}$ & log $\rm{M_{*}}$ & $\rm{\chi}^{2}_{\nu}$\\
 &  &  ($\mu$Jy) & ($\mu$Jy)& (mJy) & (mJy) & (mJy) & (mJy)& (mJy) &(\msun)& \\
\hline 
J071433.54$-$363552* & 0.88 & - & $7.64 \pm 0.44$ &< 0.012 & $0.039 \pm 0.011 $ &$0.99 \pm 0.12$ &$4.01 \pm 0.84$& $2.4\pm0.3$& 9.97 &1.06\\
J071912.58$-$334944 & 1.63 & - & $12.0 \pm 0.67$ &< 0.011 & $0.081 \pm 0.012$ &$1.93 \pm 0.12$ &$4.06 \pm 0.88$ & $5.2\pm0.6$&    10.83           &4.28\\
J081131.61$-$222522 & 1.11  &- & $24.4 \pm 1.42$ &$0.132\pm 0.0086$ &$0.611\pm0.023$ & $5.62\pm0.17$& $7.61\pm1.17$& < 1.8&      10.90* & 9.71        \\
J082311.24$-$062408 & 1.75 &  - & $13.9 \pm 0.49$ &$0.118 \pm 0.0078 $& $0.441\pm0.019$& $4.08\pm0.15$& $10.42\pm0.97$& < 1.8&   11.07*            &16.8\\
J130817.00$-$344754 & 1.65 &$19.1 \pm 0.19$ & $20.1 \pm 0.37$&$0.086\pm0.0056$ &$0.248\pm0.013$ & $3.36\pm0.12$&$9.12\pm 0.73$ & $1.38\pm0.34$&  10.66* &10.0\\
J134331.37$-$113609 & 2.49 & - & $21.1 \pm 2.03 $ &$0.024\pm0.0057$ &$0.136\pm0.013$ &$1.61\pm0.12$ & $3.81\pm0.79$& $2.34\pm0.31$ &       11.24*     &4.09 \\
J140050.13$-$291924 & 1.67& $36.5 \pm 0.19 $ & $57.4\pm 0.45$&$0.110\pm0.0063$ &$0.501\pm0.018$ & $5.58\pm0.14$& $11.85\pm0.77$ &< 0.90 & 11.02*  & 9.50\\
J141243.15$-$202011 & 1.82 &$11.2 \pm 0.17 $ & $34.5 \pm0.41$& $0.092\pm0.0063$&$0.333\pm0.015$ &$3.39\pm0.13$ &$7.41\pm0.78$ & $2.55\pm0.63$ &  11.57* &9.30\\
J143419.59$-$023543 & 1.92 & - & $18.2 \pm 0.39 $ &$0.058\pm0.0056 $ &$0.257\pm0.014$ &$2.13\pm0.11$ &$5.04\pm0.71$ & < 0.9&   11.21* &10.7          \\
J143931.76$-$372523 & 1.19 & - & $34.3 \pm 0.42 $ &$0.027\pm0.0071$ & $0.115\pm0.013$& $2.34\pm0.12$&$3.92\pm0.83$& < 0.6&  10.24* &2.77            \\
J150048.73$-$064939 & 1.50 &  $11.1 \pm 0.16$ & $22.6 \pm0.49$&$0.068\pm0.0065$&$0.293\pm0.016$ & $6.26\pm0.17$&$15.77\pm0.94$& $6.11\pm0.28$ &11.12*&5.91  \\
J151003.71$-$220311 & 0.95 &$11.7 \pm 0.16 $ & $37.1 \pm1.57$&$0.143\pm0.0095$&$0.411\pm0.020$ & $5.34\pm0.18$&$14.87\pm1.09$& < 0.9 & 10.54*  & 7.09\\
J151310.42$-$221004 & 2.20 & $9.03 \pm 0.13 $ &$27.8\pm0.46$ &$0.037\pm0.0082$&$0.214\pm0.018$ &$2.64\pm0.16$ &$9.71\pm1.10$ & $4.86\pm0.27$&  11.74* & 4.97\\
J151424.12$-$341100 & 1.09 & $12.2 \pm 0.15 $ & $26.6 \pm0.43$&$0.076\pm0.0091$&$0.184\pm0.019$ &$3.21\pm0.16$ &$7.01\pm1.03$ & < 0.9&10.53 & 1.51  \\
J152116.59+001755 & 2.63   &   - & $18.8 \pm 0.43 $ &$0.039\pm0.0046$&$0.274\pm0.014$ &$5.51\pm0.15$ &$9.51\pm0.70$& $1.19\pm0.28$&   11.64 &12.4  \\
J154141.64$-$114409 & 1.58  & - & $12.4 \pm 0.41 $ &$0.032\pm0.0077$ & $0.155\pm0.017$&$2.91\pm0.16$ &$10.74\pm1.14$&$1.2\pm0.3$ &   10.89* &2.68\\
J163426.87$-$172139 & 2.08  &$4.26 \pm 0.15$ & - & $0.039\pm0.0094$ &$0.101\pm0.018$ &$1.70\pm0.17$ & $3.57\pm1.15$ & < 0.84&   11.08*  &2.17\\
J164107.22$-$054827 & 1.84 & - & $25.2 \pm 0.44 $ &$0.086\pm0.0083$ & $0.423\pm0.020$&$3.14\pm0.15$ & $6.26\pm0.89$ & $2.3\pm0.29$&   11.29&12.7   \\
J165305.40$-$010230 & 2.02 &- & $21.3 \pm 0.53 $ &$0.083\pm0.0074$ & $0.191\pm0.015$& $2.56\pm0.14$&$5.31\pm0.93$ & < 0.78& 10.92  & 7.88          \\
J165742.88$-$174049* & 2.00*& - & $22.2 \pm 0.49 $ & $0.073\pm0.0102$&$0.186\pm0.026$ & $2.82\pm0.24$& $8.60\pm1.01$ & < 0.78&   10.86  &3.06\\
J170204.65$-$081108 & 2.82 &  - & $8.13 \pm 0.37 $ &$0.021\pm0.0690$ &$0.074\pm0.053$ &$3.05\pm0.26$ & $12.32\pm1.40$ & < 1.02&     11.23* &  2.84    \\
J170325.05$-$051742 & 1.80 &$3.61 \pm 0.15$ & $8.94 \pm0.45$& $0.021\pm0.0082$& $0.199\pm0.018$& $2.35\pm0.24$&$11.66\pm1.42$& $1.02\pm0.27$&  10.76& 3.14  \\
J170746.08$-$093916* & 2.00*& $11.4 \pm 0.18$ & $52.4 \pm0.53$&$0.119\pm0.0073$&$0.342\pm0.020$ & $3.46\pm0.28$& $3.27\pm1.26$ & < 1.02&  11.99&5.45 \\
J193622.58$-$335420 & 2.24 & $7.93 \pm 0.17 $ & $14.7 \pm0.45$&$0.031\pm0.0069$ &$0.127\pm0.016$ & $2.34\pm0.14$& $5.27\pm0.96$ & $1.86\pm0.36$& 10.95 &5.11\\
J195141.22$-$042024 & 1.58 & - & $24.1 \pm 0.58$  & $0.030\pm0.0178$ &$0.065\pm0.036$ & $2.55\pm0.15$& $8.56\pm1.02$ &< 1.03&  11.42* &1.60          \\
J195801.72$-$074609 & 1.80 & $5.66 \pm 0.17 $ &$16.1\pm 0.44$&$0.056\pm0.0086$&$0.203\pm0.018$ & $3.29\pm0.16$& $7.44\pm1.06$ & < 0.93&  11.01* &7.82 \\
J200048.58$-$280251 & 2.28 & - & $12.9 \pm 0.45 $ &$0.027\pm0.0169$&$0.113\pm0.017$ & $3.21\pm0.17$& $7.19\pm1.20$ & < 0.96&10.81& 7.14    \\
J202148.06$-$261159 & 2.44 &$1.09 \pm 0.12 $ & $6.89 \pm0.50$&<0.015 & <0.065 & $10.3\pm0.15$ & $6.27\pm1.01$ & $4.4\pm0.38$&  11.52&1.35 \\
J204049.51$-$390400 & 2.00* &$1.84 \pm 0.16 $ & $9.56\pm0.57$&$0.070\pm0.0077$ &$0.254\pm0.017$ & $2.75\pm0.15$& $4.02\pm0.91$ & $5.1\pm0.43$&  12.14&8.95 \\
J205946.93$-$354134 & 2.38 &- & $18.8 \pm 0.55 $ &  $0.052\pm0.0069$& $0.182\pm0.0015$&$2.94\pm0.014$ &$4.74\pm0.99$& < 0.99&   11.04 &8.98  \\
\hline
\end{tabular}
\end{table}
\end{landscape}

\section{Spectral properties}
\label{sec:specprop}
\subsection{NIR arm detections}
\label{nirprop}
Although spectra were taken with all three arms of X-shooter, here we focus on the near-IR results, where with source redshifts between $z \sim 1-3$, \halpha{}, \hbeta{} and \oall{} are found in the range of this arm. A brief discussion of the UVB/VIS data is presented in Section \ref{sec:otherprop}.
\par
We detected emission lines in 18 out of the 27 sources observed with X-shooter. Out of the 9 sources without detected IR emission lines, five had no discernible IR continuum in their spectra. These five sources were only observed in the ISAAC $K_{s}$ band, which may have affected the quality of the positional information available and therefore implies that for these sources the slit probably missed the target. The remaining four sources showed measurable continuum, so it is likely that here the targeting was correct. The detected continuum in two of these sources was relatively weak compared to others with IR detections, but the two sources with stronger continuum had fluxes that agreed within error to the magnitudes from the ISAAC photometry. It is therefore likely that in these cases the emission lines were too faint to detect or just so heavily reddened by dust as to be indistinguishable from the underlying continuum. Other reasons for these undetected lines could include the emission lines falling in regions of strong sky absorption, simply being too faint, or a misplaced slit that missed the target. It is uncertain which, if any, of these factors were responsible for non-detections.
\par

\begin{figure*}
\centering
\includegraphics[width = 0.9\textwidth]{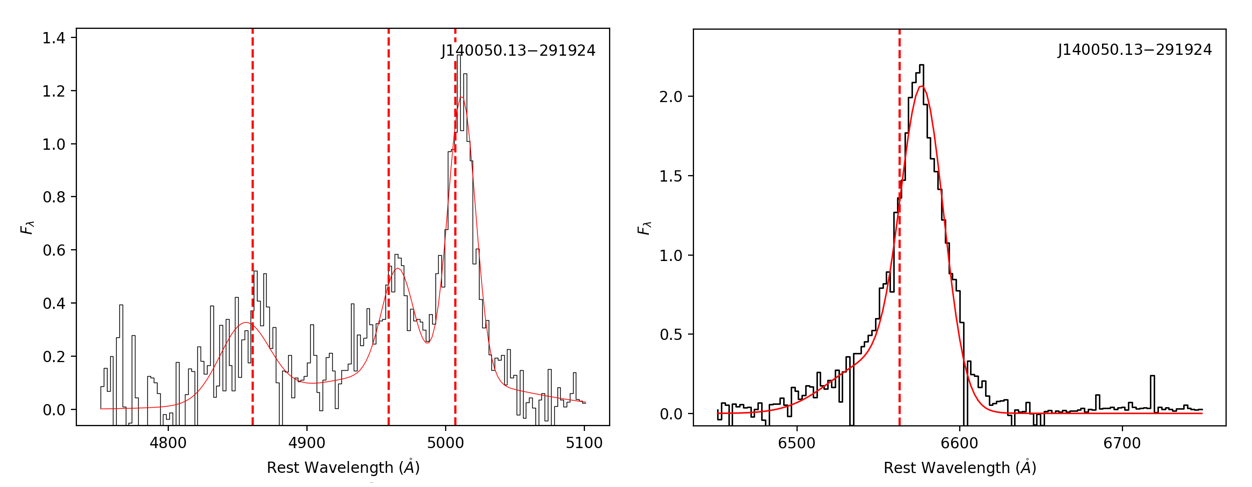}
\caption{\label{fig:spec} Rest frame X-shooter spectra in the region surrounding \hbeta{}, \oall{} (left) and \halpha{} (right) for WISE J140050.13$-$291924. Flux densities are in the units $10^{-16}$ $\rm{erg\,s^{-1}cm^{-2}\ang{}^{-1}}$. Reduced spectra and fitted model are represented by black histogram, and red lines, respectively. For \hbeta{} and \oall{} spectra, the red line represents a combined model of the multi-Gaussian model fitted to each individual peak. For the \halpha{} spectra the red line also represents a multi-Gaussian model fitted to the blended \halpha{} and \nii{} profile. Vertical dashed red lines indicate the rest frame wavelengths of the \hbeta{}, \oall{} and \halpha{} emission lines based on the redshifts from \citet{Lonsdale2015}. }  
\end{figure*}

The 18 IR line-detected sources discussed further below have at least one detected \halpha{}, \hbeta{}, or \ofive{} emission line. Following the recipe described by \citet{Greene2005}, a multi-Gaussian model was used to fit the 1D spectra to obtain basic properties. This was required due to the asymmetric nature of the emission lines, with both broad and narrow components being separately fitted. The \halpha{} emission line (including the \nii{} doublet) was modelled independently, while \hbeta{} and the \oall{} doublet models have been combined into a compound model. While the \halpha{} emission line could not be resolved from the \nii{} doublet, we assume the \halpha{} to be the dominant profile for both the measured flux and FWHM. For the \hbeta{} and \oall{} emission lines, each peak was modelled individually, with their models then combined and the separation between the three peaks fixed to their known wavelength differences. The \ofour{} emission line was solely modelled to provide a systematic check on our measurements of the \ofive{} line, with the peak ratio of \ofive{} to \ofour{} constrained to 2.88 \citep{Kim2013}, and their modelled widths fixed to be the same value.The addition of the \ofour{} peak to our models helped to accurately distinguish the extent of the \ofive{} emission line from the underlying noise allowing a systematic check on our values measured. Note that all the \ofive{} values in Table \ref{tab:spec} are determined from the constrained \ofive{} model while a combined model of all the lines (Figs. \ref{fig:spec} and \ref{fig:specap}) is plotted for convenience.

\begin{figure*}
\centering
\includegraphics[width =\textwidth]{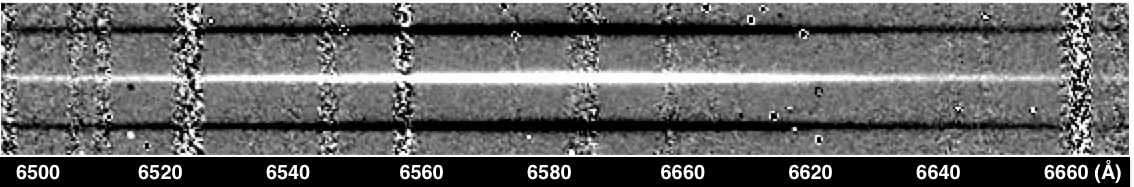}
\caption{\label{fig:2d} The 2D spectrum of WISE J140050.13$-$291924 centred on the \halpha{} emission line. This subsection spans 15.75'' in the spatial direction and 200\,\ang{} in the wavelength direction. The same source is shown in Fig \ref{fig:2d}.}
\end{figure*}

Following previous studies \citep[eg.][]{Greene2005,Assef2011,Jun17,Wu2018,hotdog2020}, the minimum number of Gaussian components needed to obtain an acceptable fit as defined below, are used to represent the final emission line. This meant that the final model of the emission lines was composed of as many individual models needed for the final model to provide a low $\chi^{2}$ fit to the binned data. Compared to other estimates \citep{Kim2013}, a multi-Gaussian approach allowed for both narrow and broad features to be included in the final compound model of each emission line. This fitting was done by limiting the amplitude and width of each Gaussian to fit the individual features of the spectra before combining into a compound model. A $\rm{\chi^{2}_{\nu}}$ analysis was performed, with additional Gaussians added into the compound model until an acceptable fit was achieved (based on $\rm{\chi^{2}_{\nu}}$), in other words; stopping when increasing the number of Gaussians generated a poorer $\rm{\chi^{2}_{\nu}}$ fit. In addition, using the 2D spectra (e.g. Fig. \ref{fig:2d}) allowed additional emission lines to be clearly detected, where they appeared blended in the 1D spectra. A Monte-Carlo approach was adopted \citep{Assef2011,Assef2015,Wu2018} to estimate the errors on the measured fluxes and line widths. Using the pipeline given spectral errors, we simulate 10,000 random spectra producing a normal-like distribution of fluxes and line widths. Standard deviations of this distribution were calculated to give flux and line width errors for each source.

\par
The X-shooter spectra shown in Fig. \ref{fig:spec} covers the rest-frame 4800 - 5500\ang{} and 6500 - 6750\,\ang{} wavelength range for source WISE J140050.13$-$291924 to demonstrate the final fitting procedure. Spectra for all 18 sources with detected emission lines are shown in Appendix \ref{sec:app}. The binned spectra are shown in black with the fitted model over-plotted in red. The spectral properties measured for the detected emission lines, including the emission line fluxes and Full Width Half Maximum (FWHM) line widths for the \halpha{}, \hbeta{} and \ofive{} detections are listed in Table \ref{tab:spec}. Note that not all emission lines were detected for each source; 13 had detected \halpha{} lines, six had detected \hbeta{} lines and 15 had detected \ofive{} lines; the exact breakdown for specific sources is listed in Table \ref{tab:spec}. Non-detections, shown by spectra in Appendix \ref{sec:appbad}, are likely to be due to the intrinsic faintness of the emission lines with the searched for emission lines likely to be lurking beneath the higher noise in the spectra. However, the luminosity of the undetected emission lines can be limited using measured properties from the other lines (see Section \ref{sec:extinc}).
\par
The spectral properties of the \ofive{}, \halpha{} and \hbeta{} emission lines are listed in Table \ref{tab:spec}. Their luminosities are subsequently used to measure the BH and host galaxy masses of our sources.

\begin{table*}
\small
\centering
\caption{\label{tab:spec} The line flux and widths measured for the 18 sources with detected emission lines.}
\begin{tabular}{l l l l l l l l}
\hline
\hline
Source ID & WISE designation & \multicolumn{3}{l}{Flux ($10^{-16}$ $\rm{erg\,s^{-1}cm^{-2}}$)} & \multicolumn{3}{l}{FWHM ($\rm{kms^{-1}}$)} \\
& & $f_{\rm{H_{\rm{\alpha}}}}$ & $f_{\rm{H_{\rm{\beta}}}}$ & $f_{\rm{[OIII]\lambda5007}}$ &\halpha{} & \hbeta{} & \ofive{} \\
\hline
1&J081131.61$-$222522 &$8.49\pm0.46$ &- & $14.76\pm0.54$ &$1440\pm150$ & -& $1260\pm140$  \\
2&J082311.24$-$062408 & $16.70\pm3.05$ & $6.04\pm0.21$ & $8.58\pm0.30$ & $2890\pm800$ & $1880\pm210$& $1090\pm510$\\
3&J130817.00$-$344754 & $28.07\pm3.70$ & $5.42\pm0.12$ & $48.00\pm3.55$ &$1640\pm80$ & $1030\pm110$& $1140\pm130$\\
4&J134331.37$-$113609 &- &- &$18.72\pm7.01$ &- &- & $2020\pm560$ \\
5&J140050.13$-$291924 & $86.07\pm4.13$ & $8.59\pm0.26$ & $32.13\pm9.32$ & $2130\pm30$ & $1200\pm260$ & $1620\pm50$ \\
6&J141243.15$-$202011 & $6.74\pm0.34 $ &-& $2.83\pm0.30$ & $2180\pm880$ &-& $670\pm270$ \\
7&J143419.59$-$023543 & $6.86\pm0.10$ &- & $14.32\pm2.07$ & $2220\pm340$ &-& $780\pm80$ \\
8&J143931.76$-$372523 & $9.48\pm0.15$ & -&- & $2680\pm280$ & -& -\\
9&J150048.73$-$064939 & $64.49\pm2.12$ &$3.66\pm0.15$& $2.35\pm0.22$ & $2350\pm90$ &$2230\pm150$ &$1680\pm520$\\
10&J151003.71$-$220311 & $12.65\pm6.97$ & -& $21.11\pm1.23$& $1460\pm220$ &- &$480\pm30$ \\
11&J151310.42$-$221004 & $17.51\pm1.53$ &- &-& $3690\pm390$ &-&- \\
12&J151424.12$-$341100 &- &- & $13.86\pm5.22$ &- & -& $1040\pm320$ \\
13&J154141.64$-$114409 & $23.35\pm3.92$ &-& $14.77\pm1.86$ & $2430\pm470$ &- & $1390\pm280$\\
14&J163426.87$-$172139 & $15.00\pm3.91$ &-& $5.69\pm0.14$& $2680\pm260$ &- & $1350\pm200$\\
15&J170204.65$-$081108 &- &- & $13.86\pm7.21$ & -& -& $1570\pm500$\\
16&J195141.22$-$042024 & $55.86\pm 6.52$ & $10.27\pm3.96$ &$76.85\pm0.16$ & $1430\pm390$ & $1060\pm120$ & $1190\pm30$\\
17&J200048.58$-$280251 & - & - &$5.56\pm0.14$& - & - & $1690\pm270$ \\
18&J204049.51$-$390400 & - & $3.30\pm0.82$  & - &- & $490\pm70$ & -\\
\hline
\end{tabular}
\end{table*}

\subsection{UVB + VIS arm detections}
\label{sec:otherprop}

Alongside the near-IR lines used to yield AGN properties, we also explored the data from the other two X-shooter arms; UVB and VIS, spanning wavelength ranges 300 - 559.5\,nm and 559.5 - 1024\,nm, respectively. Searching for two common emission lines, \lya{} and \civ{}, we aim to confirm the consistency of our non-detections of emission lines and/or continuum in the near-IR spectra, and investigate any other emission lines in parallel. 

\begin{figure}
\centering
\includegraphics[width = 0.48\textwidth]{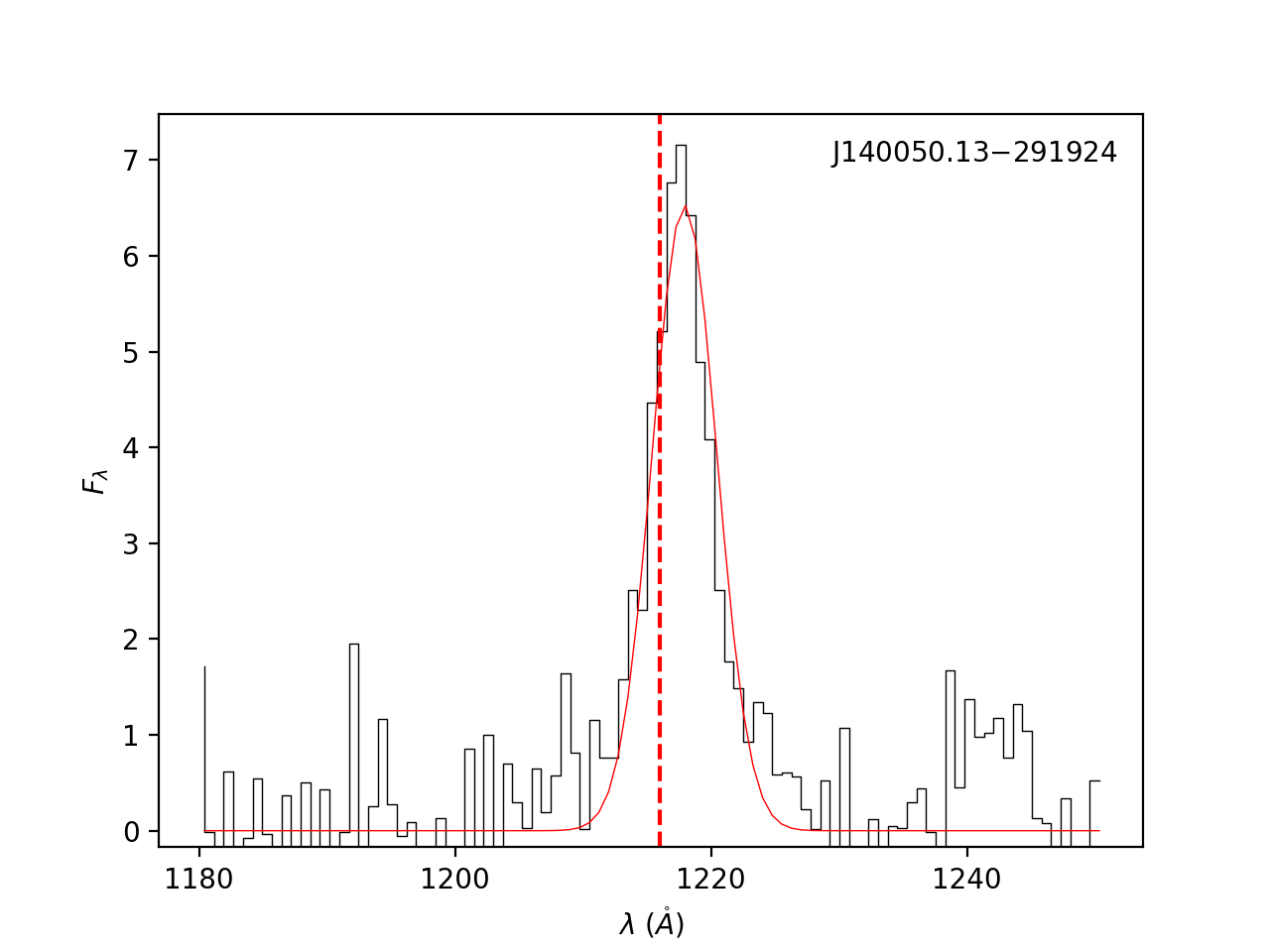}
\caption{\label{fig:lya} Rest frame wavelength X-shooter spectra in the region surrounding the \lya{} emission line for WISE J140050.13$-$291924. Flux density is in the units $10^{-16}$ $\rm{erg\,s^{-1}cm^{-2}\ang{}^{-1}}$. Reduced spectra and fitted model are represented by black histogram and red line respectively. The vertical dashed red line indicates the rest frame wavelength of the \lya{} emission line based on the redshift from \citet{Lonsdale2015}.}
\end{figure}

Out of the 27 sources with spectra from the UVB arm, four sources had a redshifts too low for \lya{} to be observed. A further two sources had redshifts where the spectra was too noisy at the expected \lya{} wavelength ($\rm{z \leq 1.6}$) to clearly detect an emission line. This was due to being near the end of the wavelength range covered. Out of the remaining 21 sources, we detect 10 sources with \lya{} emission, all of which have detected near-IR emission lines as well. By fitting a simple single Gaussian model, we measure the flux and FWHM of the 10 detected emission lines (see Table \ref{tab:uvbspec}), with errors found using the same Monte Carlo approach discussed above. Fig. \ref{fig:lya} shows an example detection of the \lya{} line for WISE J140050.13$-$291924.

We checked for evidence of velocity offsets of the \lya{} emission line by comparing the centre of our fitted Gaussian emission line with the 1216\,\ang{} position that we expect the line to be centred on. To account for errors in measuring each source's redshift, we used a minimum wavelength difference of $2.43\,\ang$, equivalent to $\Delta z \sim 0.01$ or a shift of $\sim 600\,\rm{kms^{-1}}$, to determine a velocity offset that could be indicative of other processes. Out of the 10 sources with \lya{} detentions, three sources had been redshifted with a mean offset of $\sim 6.8 \ang$ ($\sim 1680 \rm{kms^{-1}} $), while a fourth was blueshifted by 8.7\,\ang{} ($\sim 2150 \rm{kms^{-1}} $) in comparison to the expected 1216\,\ang{} position. These four velocity offsets could either indicate possible absorption by the surroundings or be evidence for potential outflows. There appears to be no common velocity offset across the 10 sources with detections of the \lya{} line, suggesting other rest optical lines may be more reliable to extract results from.

\begin{figure}
\centering
\includegraphics[width = 0.48\textwidth]{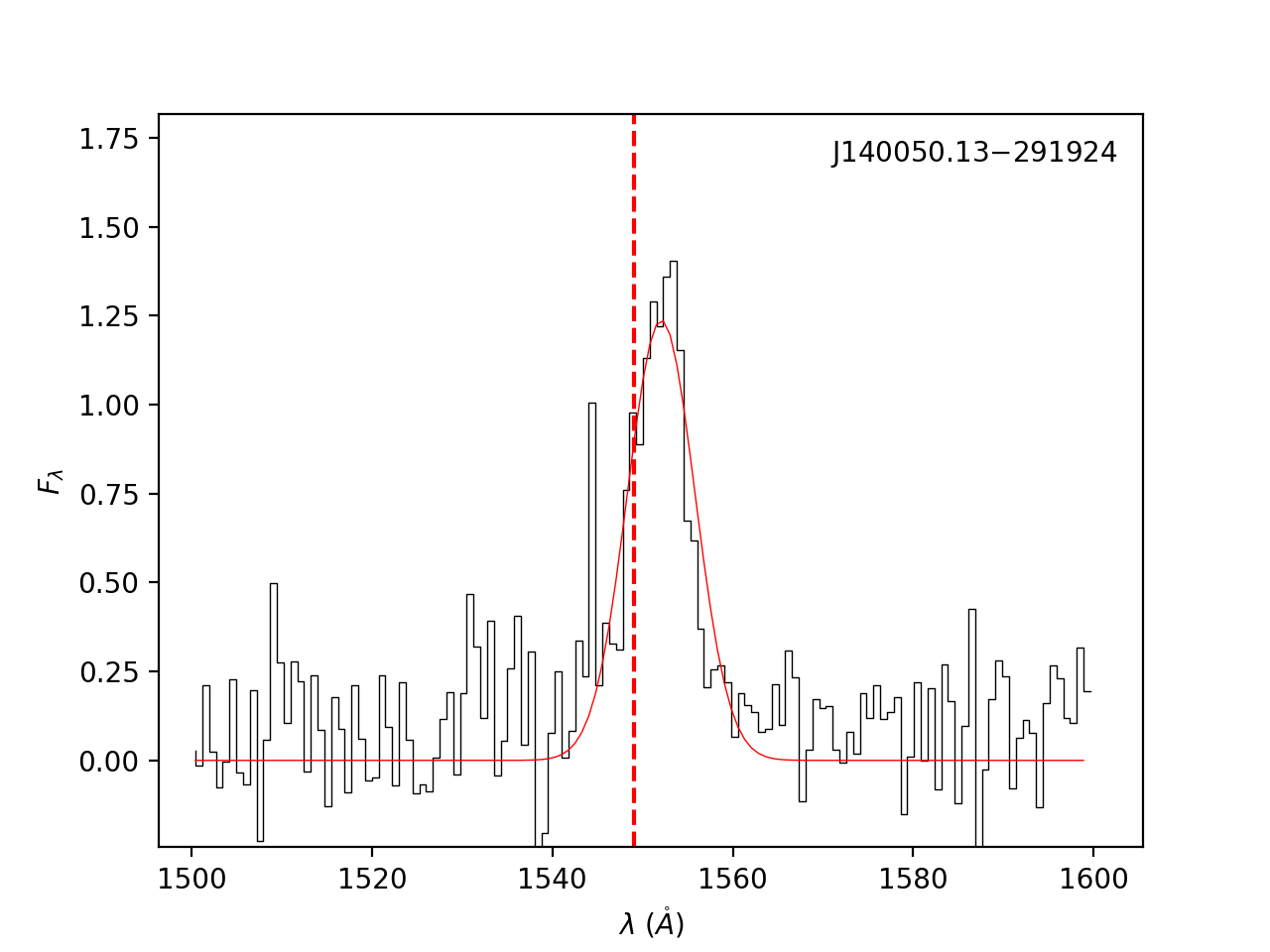}
\caption{\label{fig:1400civ} Rest frame X-shooter spectra in the region surrounding the \civ{} emission line for WISE J140050.13$-$291924. Flux density is in the units $10^{-16}$ $\rm{erg\,s^{-1}cm^{-2}\ang{}^{-1}}$. Reduced spectra and fitted model are represented by black histogram and red line respectively. The vertical dashed red line indicates the rest frame wavelength of the \civ{} emission lines based on the redshift from \citet{Lonsdale2015}. }
\end{figure}

In addition to \lya{}, we searched for the typically fainter \civ{} line. Though the \civ{} line is a doublet we are measuring it as a single emission line at a rest frame wavelength of 1549\,\ang{}, where one source had a redshift too low for it to be observed. For two of our sources the \civ{} line fell in the wavelength range of the VIS arm instead of the UVB arm, but no detections of this emission line were made for either of these sources, even with a lack of noise in the spectra. In total we detect five \civ{} lines in the UVB arm of X-shooter, all of which have corresponding \lya{} and therefore IR emission line detections. Again, their line fluxes and widths are measured by fitting a single Gaussian model, with results given in Table \ref{tab:uvbspec} and an example fit given in Fig. \ref{fig:1400civ}.

\begin{figure}
\centering
\includegraphics[width = 0.48\textwidth]{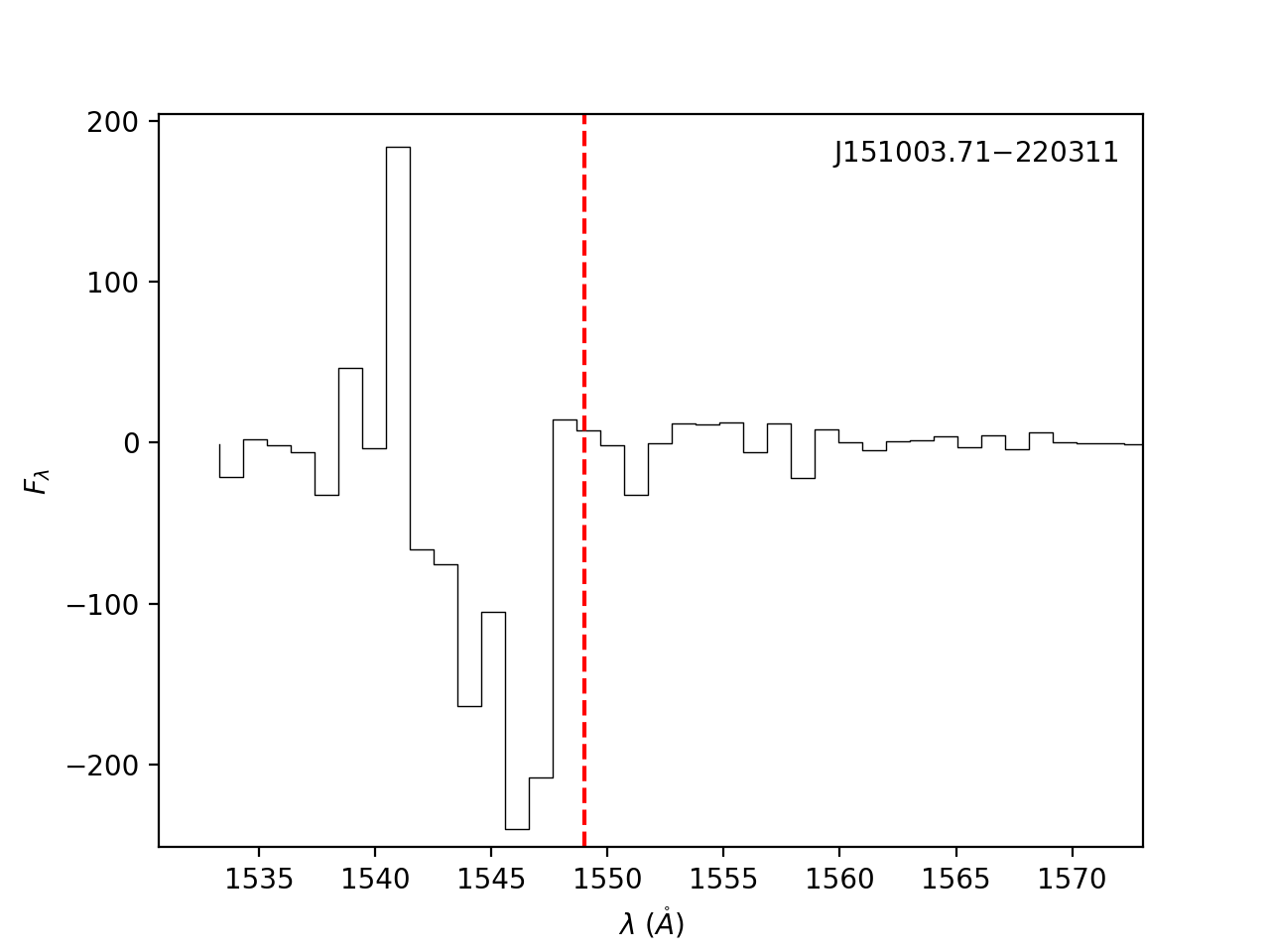}
\caption{\label{fig:abs} Rest frame X-shooter spectra in the region around the \civ{} emission line rest wavelength (expected at 1549 \ang{}) for the source WISE J151003.71$-$220311, demonstrating a clear absorption line, unique to our sample. Flux density is in the units $10^{-16}$ $\rm{erg\,s^{-1}cm^{-2}\ang{}^{-1}}$. The vertical dashed red line indicates the rest frame wavelength of the \civ{} absorption line based on the redshift from \citet{Lonsdale2015}. }
\end{figure}

We also checked for any continuum emission visible in the UVB spectra. For the 8 sources with near-IR detections but without \lya{} (or \civ{}) detections, only one source had an obvious continuum, J151003.71$-$220311. This exception had both a strong continuum and a narrow absorption line at the \civ{} wavelength, shown in Fig. \ref{fig:abs}. As we had near-IR line detections for these sources, and therefore did not miss our source with the slit, it is unclear why we have no detections of these lines as there was also no overlap with common atmospheric lines which may have affected our detections. As an additional check, we stacked these 8 spectra, but still found no continuum or emission lines.

In total we detect 15 emission lines from the UVB arm across 10 different sources. All of these sources have corresponding IR detections, meaning that those five sources with neither emission line detections, nor continuum, were likely missed and not positioned correctly on the slit. The \lya{} emission lines we detect are the brightest of all detected lines with a mean luminosity of $\rm{L \sim 10^{44.8}\, erg\,s^{-1}}$. They are $\sim 2.8 \times$ brighter than the \civ{} lines, but the \lya{} emission lines are relatively narrow compared to the broad \civ{} and IR lines.

\begin{table*}
\small
\centering
\caption{\label{tab:uvbspec} The line flux and widths measured for the 10 sources with detected emission lines from the UVB arm.}
\begin{tabular}{l l l l l l l l}
\hline
\hline
Source ID & WISE designation & \multicolumn{2}{l}{Flux ($10^{-16}$ $\rm{erg\,s^{-1}cm^{-2}}$)} & \multicolumn{2}{l}{log Luminosity ($\rm{erg\,s^{-1}}$)} & \multicolumn{2}{l}{FWHM ($\rm{kms^{-1}}$)} \\
& & $f_{\rm{Ly\alpha}}$ & $f_{\rm{\civ{}}}$ & $\rm{L}_{\rm{Ly\alpha}}$ & $\rm{L}_{\rm{\civ{}}}$ & \lya{} & $\rm{\civ{}}$  \\
\hline

2&J082311.24$-$062408 & $62.00\pm3.02$  & $24.14\pm0.85$ &  $44.12\pm0.02$ & $43.70\pm0.02$ & $2030\pm150$ & $2480\pm130$ \\
3&J130817.00$-$344754 & $26.16\pm2.83$  & $10.10\pm0.90$ &  $43.67\pm0.05$ & $43.26\pm0.04$ & $2080\pm350$ & $4360\pm560$ \\
4&J134331.37$-$113609 & $20.87\pm4.82$  & $11.28\pm0.27$ &  $44.01\pm0.01$ & $43.74\pm0.01$ & $1750\pm50 $ & $2170\pm60 $ \\
5&J140050.13$-$291924 & $41.18\pm2.19$  & $11.64\pm0.534$ &  $43.87\pm0.02$ & $43.33\pm0.02$ & $1460\pm110$ & $2180\pm160$ \\
6&J141243.15$-$202011 & $36.99\pm2.60$  & -                &  $43.92\pm0.03$ & -                & $3330\pm310$ & -            \\
7&J143419.59$-$023543 & $16.45\pm1.31$  & -                &  $43.63\pm0.03$ & -                & $3560\pm350$ & -            \\
10&J150048.73$-$064839& $246.41\pm24.11$& -                &  $44.54\pm0.04$ & -                & $1380\pm50 $ & -            \\
11&J151310.42$-$221004& $5.09\pm1.68$   & -                &  $43.7\pm0.12$ & -                & $2350\pm790$ & -            \\
13&J154141.64$-$114409& $29.18\pm1.55$  & $7.48\pm1.13 $ &  $43.62\pm0.03$ & $43.08\pm0.06$ & $2250\pm310$ & $2150\pm610$ \\
17&J200048.58$-$280251& $3.59\pm0.39$   & -                &  $43.14\pm0.05$ & -                & $1180\pm150$ & -            \\

\hline
\end{tabular}
\end{table*}

\subsection{Line properties}

We have discussed the spectral properties of the emission lines found in the NIR and UVB arms. We detect a total of 18 emission lines out of the 27 sources with X-shooter spectra, with 15 \ofive{}, 13 \halpha{} and six \hbeta{} detentions with the NIR arm, and 10 \lya{} and five \civ{} detections in the UVB arm. We however note for the UVB arm there was a redshift factor preventing the detection of certain lines. We had three sources with redshifts too low to observe \lya{}, and for a further four sources the \lya{} line was in a region of high noise near the end of the arm. Similarly one source was not observed at the \civ{} wavelength, and while three had redshifts such that \civ{} should have been observed with the VIS arm, we measure no detections. 

There is a stark contrast between the number of detections for each emission line, with 15 \ofive{} detections and only four \civ{} detections. Whilst the number of detections is likely to depend on the X-shooter arm used (NIR or UVB) and the rest wavelength of each emission line, there is a clear trend of a higher number of detections for the brighter emission lines. With luminosities $\sim 3.6\, \times$ less than \halpha{}, we have half as many detections of the \hbeta{} line. Similarly for the UVB line detections, we detect twice as many \lya{} lines compared with \civ{} with a mean luminosity difference of a factor of $\sim 2.2$. However, with comparable mean luminosities of $\rm{\sim 10^{43.60}\,erg\,s^{-1}}$ and $\rm{\sim 10^{43.68}\,erg\,s^{-1}}$, we have 15 emission line measurements of \ofive{} and 13 \halpha{}. While \lya{} is the brightest of all the detected lines, it is clear that our lack of detections of the \hbeta{} and \civ{} lines are likely due to their intrinsic faintness.

An additional similarity across some of the emission lines is their asymmetric appearance. Most notably for the \halpha{} emission lines of J130817.00$-$344754, J150048.73$-$064839 and J195141.22$-$042024 (see Fig. \ref{fig:specap}), but also the \lya{} emission line of J130817.00$-$344754, indicating the possibility of some spatially extended structure. We also find evidence of an underlying broad red component in our \lya{} detections by stacking them together, again indicating a more unusual morphology compared to regular AGN.


\section{Black hole and host galaxy masses}
\label{sec:bh}

With the availability of line luminosities and widths for three different emission lines from the near-IR spectra, we derive two different methods to independently determine the black hole masses of our sources. First using the \ofive{} and then the Balmer emission lines, we present three independent estimates for the black hole and host galaxy masses of our sources, along with a comparison to SED derived masses by \citet{Lonsdale2015} and spectra derived masses from \citet{Kim2013}.

Throughout this work we assume that our detected emission lines are broadened by the gravity of the SMBH as is true in typical quasars. However it has been recently suggested that for the similar galaxy population, HotDOGs, the Balmer emission lines are actually broadened by contamination from outflows \citep{hotdog2020}, along with the \ofive{} emission line of some obscured AGN \citep{zakamska16}. This could have an effect on our measured widths and therefore our derived masses, however both of the populations have substantial differences compared to our radio AGNs.

\subsection{\ofive{}}
\label{sec:oiii}

\begin{table*}
\small
\centering
\caption{\label{tab:mass}The calculated luminosity ($\rm{erg\,s^{-1}}$) and range of BH and host galaxy masses within  1$\rm{\sigma}$ error, (given in solar masses) for those sources with measured \ofive{} emission lines. The $\rm{M_{BH}/M_{*}}$ relation is given using the $\rm{M_{*}}$ from Table \ref{tab:phot}. See Section \ref{sec:oiii} for full details on calculations.}
\begin{tabular}{l l l l l l}
\hline
\hline
Source ID & WISE designation & log {$\rm{L}_{\text{[OIII]}\rm{\lambda}5007}$} ($\rm{erg\,s^{-1}}$) & $\log\left(\rm{\frac{{M}_{BH}}{\text{M}_{\odot}}}\right)$ & $\log\left(\rm{\frac{{M}_{Host}}{\text{M}_{\odot}}}\right)$ & $\rm{M_{BH}/M_{*}}$ ($10^{-3}$)\\
\hline
1&J081131.61$-$222522 & $43.00\pm1.44$ & 8.39 - 8.42 & 10.86 - 11.14 & $  3.19 \pm0.12 $\\
2&J082311.24$-$062408 & $43.25\pm1.45$ & 8.64 - 8.67 & 10.90 - 11.26 & $  3.81 \pm0.14 $\\
3&J130817.00$-$344754 & $43.94\pm1.13$ & 9.31 - 9.37 & 11.60 - 11.98 & $ 47.73\pm 3.67$\\
4&J134331.37$-$113609 & $43.96\pm0.42$ & 9.16 - 9.51 & 11.36 - 11.93 & $ 12.37\pm 6.07$\\
5&J140050.13$-$291924 & $43.77\pm0.54$ & 9.02 - 9.28 & 11.31 - 11.89 & $ 13.45\pm 4.68$\\
6&J141243.15$-$202011 & $43.80\pm0.97$ & 8.16 - 8.25 & 10.40 - 10.83 & $  0.43\pm 0.4$\\
7&J143419.59$-$023543 & $43.57\pm0.84$ & 8.91 - 9.04 & 11.11 - 11.58 & $  5.83\pm 0.91$\\
9&J150048.73$-$064939 & $42.53\pm1.03$ & 7.89 - 7.97 & 10.23 - 11.61 & $  0.64\pm 0.06$\\
10&J151003.71$-$220311 & $42.99\pm1.24$& 8.37 - 8.42 & 10.89 - 11.18 & $  7.02\pm 0.42$\\
12&J151424.12$-$341100 & $42.95\pm0.42$& 8.15 - 8.50 & 10.63 - 11.23 & $  0.38\pm 0.18$\\
13&J154141.64$-$114409 & $43.38\pm0.90$& 8.72 - 8.83 & 11.04 - 11.46 & $ 7.73\pm 1.04$\\
14&J163426.87$-$172139 & $43.25\pm1.60$& 8.65 - 8.67 & 10.79 - 11.18 & $  3.80\pm 0.97$\\
15&J170204.65$-$081108 & $43.98\pm0.28$& 9.06 - 9.56 & 10.94 - 11.91 & $ 11.90\pm 0.93$\\
16&J195141.22$-$042024 & $44.09\pm2.68$& 9.50 - 9.60 & 11.81 - 12.12 & $ 12.11\pm 0.03$\\
17&J200048.58$-$280251 & $43.12\pm0.61$& 8.40 - 8.62 & 10.47 - 11.09 & $  4.98\pm 1.43$\\
\hline
\end{tabular}
\end{table*}

We assume that the bolometric luminosity should not exceed the Eddington luminosity at any given BH mass and use the \ofive{} line luminosity as a proxy for the bolometric luminosity in order to calculate a lower limit for the BH masses of the targeted sources. With 15/27 \ofive{} emission lines detected, 15 BH and inferred host masses are presented in Table \ref{tab:mass}. Mass calculations using the \halpha{} and \hbeta{} detected emission lines are discussed in Section \ref{sec:ha} and presented in Table \ref{tab:appmass}. With a conversion factor of $\rm{L}_{Bol} = \rm{3200\, L_{[OIII]}}$ \citep{Shen2011}, a correlation derived by comparing the spectral properties of broad line quasars from SDSS DR7, Equation \ref{eq:omass} can be used to calculate the BH masses from the \ofive{} line luminosities (in $\rm{L_{\odot}}$) with a choice of Eddington ratio, $\rm{\lambda}_{Edd}$.

\begin{equation}
\frac{\rm{M}_{BH}}{\rm{M}_{\odot}} = \frac{3200 \times \rm{L_{[OIII]}}}{\rm{\lambda}_{Edd} \times 3.28\times10^{4}}
\label{eq:omass}
\end{equation}

The masses in Table \ref{tab:mass} are calculated as classical lower limits assuming $\rm{\lambda}_{Edd} = 1$. Host galaxy masses are then calculated assuming the evolution of the ${M}_{\rm{BH}}$ - ${L}_{\rm{Host}}$ relation through cosmic time \citep{Decarli10}, in terms of the parameter $\Gamma = \rm{\frac{M_{BH}}{M_{Host}}}$.

\begin{equation}
\log \Gamma = (0.28\pm0.06)z - (2.91\pm0.06)
\label{eq:hostmass}
\end{equation}

Our estimates of BH mass are inversely proportional to their assumed Eddington ratios with $\rm{M_{BH} \propto 1/{\lambda}_{Edd}}$. A choice of $\rm{\lambda}_{Edd} = 1$ will produce a lower limit value of BH mass under the assumption that the source is in hydrostatic equilibrium. However, if the source accretes at a super-Eddington rate, the BH mass required for a source of the same luminosity will be lower, as discussed further in Section \ref{sec:edd}. The choice of $\rm{\lambda}_{Edd}$ made scales the BH masses and will affect the final calculated value, whether we assume $\rm{\lambda}_{Edd} =1$ \citep{Kim2013}, 0.25 \citep{Lonsdale2015} as is typical of quasars at $z\sim2$ \citep{Kormendy13} or a super Eddington rate, $\rm{\lambda}_{Edd} \geq1$ \citep{Assef2015}.

There are other assumptions that can also introduce systematic uncertainties in our measurements of the BH masses. The assumed bolometric correction factor of 3200 may be low for these radio AGNs \citep{stern12}, causing over estimates of both the BH and host galaxy masses. In addition, uncertainties in the redshift measurements\footnote{Eg., In these calculations the redshift of J204049.51$-$390400 was taken as 2.0 since no spectroscopic redshift is available.} and the evolution of the BH--host relation can affect the accuracy of the host galaxy mass measurements. Finally, all the targets are selected to be highly obscured, and our masses, shown in Table \ref{tab:mass} have not been corrected for extinction (see Section \ref{sec:extinc})\footnote{With a mean measured $\rm{E(B-V)} = 0.87$ we would expect the BH masses to be underestimated by an 0.5 orders of magnitude without this correction.}.

\subsection{\halpha{} and \hbeta{}}
\label{sec:ha}
Following \citet{Greene2005}, the width and luminosity of the \halpha{} or \hbeta{} emission lines can also be used to calculate BH masses. A virial mass for a BH can be inferred from the broad-line region (BLR) radius and the velocity dispersion. With an empirical relation between ${R_{\rm{BLR}}}$ and the optical continuum luminosity, we can calculate $\rm{M}_{BH}$ with a single spectrum rather than repeated reverberation mapping. Using the relations derived by \citet{Kaspi2000} and more recently \citet{bentz2009}, $\rm{R_{BLR}} \propto L^{0.5}_{5100}$ where ${L}_{5100} = \rm{\lambda} \rm{L}_{\lambda}$ at ${\lambda} = 5100 \mathring A$, with the width and luminosities of the Balmer lines, we calculate our BH masses using the relations from \citet{Greene2005}:

\begin{equation}
\label{eq:mha}
\begin{aligned}
\rm{M}_{BH} = (2.0^{+0.4}_{-0.3})\times10^{6} \left(\frac{L_{\rm{H}_{\alpha}}}{10^{42}\, erg\,s^{-1}}\right)^{0.55\pm0.02}\\
\times \left(\frac{\text{FWHM}_{\rm{H}_{\alpha}}}{10^{3}\, \rm{kms^{-1}}}\right)^{2.06\pm0.06} \rm{M}_{\odot}
\end{aligned}
\end{equation}

\begin{equation}
\label{eq:mhb}
\begin{aligned}
\rm{M}_{BH} = (3.6\pm0.02)\times10^{6} \left(\frac{L_{\rm{H}_{\beta}}}{10^{42}\, erg\,s^{-1}}\right)^{0.56\pm0.02}\\
\times \left(\frac{\text{FWHM}_{\rm{H}_{\beta}}}{10^{3}\, \rm{kms^{-1}}}\right)^{2} \rm{M}_{\odot}
\end{aligned}
\end{equation}

\begin{figure}
\centering
\includegraphics[width = 0.48\textwidth]{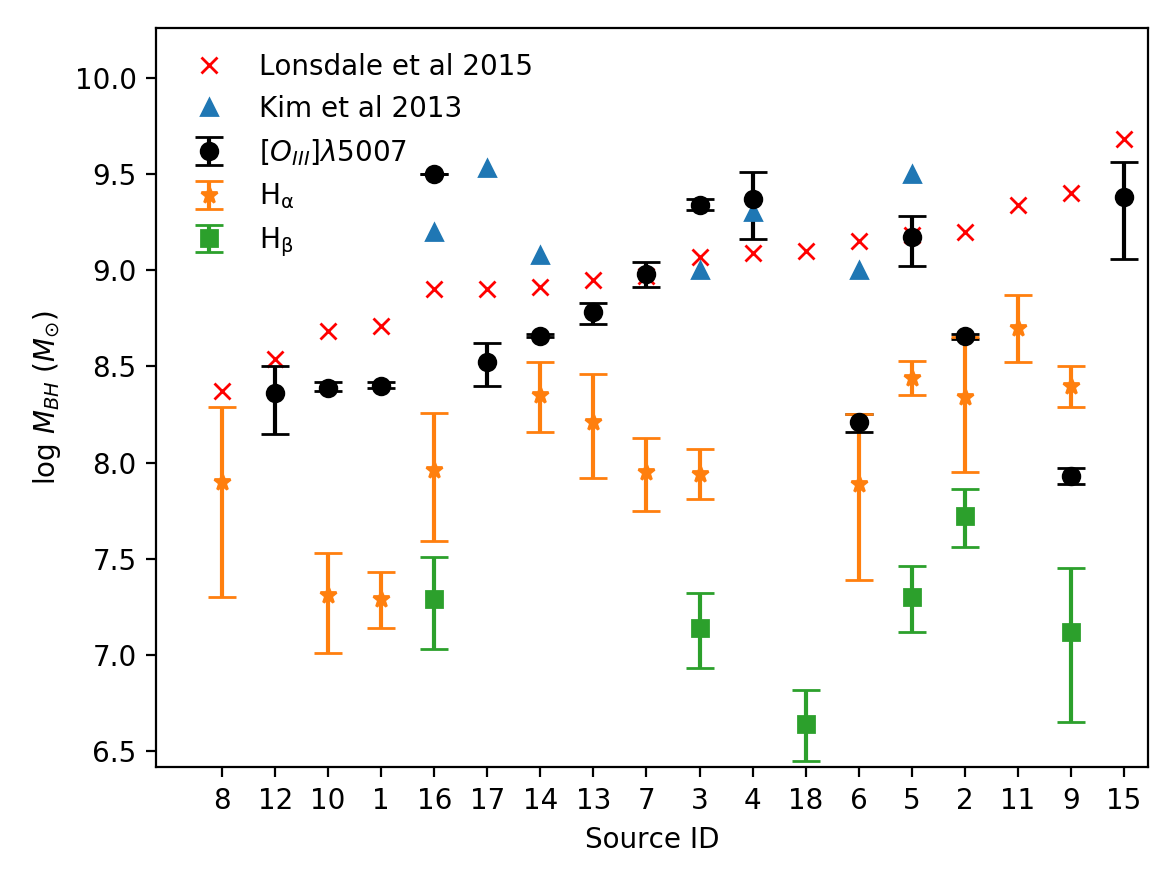}
\caption{\label{fig:bhcomp} The different BH masses calculated in this paper; using \ofive{} (black circles), \halpha{} (orange stars) and \hbeta{} (green squares) emission lines. Also included are results by other works: \citet{Kim2013} (blue triangles) and \citet{Lonsdale2015} (red crosses), where details on those measurements can be found in the original work. Plotted from smallest to largest \citet{Lonsdale2015} mass, source ID corresponds to those used in tables. Note that systematic uncertainties due to the choice of $\rm{\lambda}_{Edd}$ are not included in the errors shown, they represent the range of masses given in Table. \ref{tab:mass} and \ref{tab:appmass}.}
\end{figure}

In addition to the 15 sources with \ofive{} detections, there were 13 \halpha{} and six \hbeta{} detections, for a total of 18 sources with BH mass measurements. Using Equations \ref{eq:mha} and \ref{eq:mhb} respectively, the calculated BH and host galaxy masses are shown in Table \ref{tab:appmass} and Fig. \ref{fig:bhcomp}. 

While the \ofive{} calculated masses agree more with the previously calculated mass estimates, the \halpha{} and \hbeta{} measurements can produce estimates over an order of magnitude lower. This is likely due to the greater relative amount of extinction of \hbeta{} line compared with the \oiii{}, the high noise surrounding \hbeta{} lines, and the factor of de-blending the \halpha{} and \nii{} lines. The increased effect of extinction, along with the lower S/N, likely explains the difference between the \halpha{} and \hbeta{} calculated masses.

\begin{table*}
\caption{\label{tab:appmass} The BH and host galaxy masses (in solar masses) calculated with the detected Balmer emission lines, \halpha{} and \hbeta{}. We have also included the measured FWHMs for these sources from Table \ref{tab:spec}.}
\footnotesize
\centering
\begin{tabular}{l l l l l l l l l l l l l l}
\hline
\hline
Source ID & WISE designation & \multicolumn{2}{l}{FWHM ($\rm{kms^{-1}}$)} & \multicolumn{2}{l}{$\log\left(\rm{\frac{{M}_{BH}}{\text{M}_{\odot}}}\right)$} & \multicolumn{2}{l}{$\log\left(\rm{\frac{{M}_{Host}}{\text{M}_{\odot}}}\right)$} & $\rm{M_{BH}/M_{*}}$ ($10^{-3}$)\\  
& & \halpha{} & \hbeta{} & \halpha{} & \hbeta{}& \halpha{} & \hbeta{} & \halpha{} & \hbeta{} \\
\hline
1&J081131.61$-$222522  & $1440\pm150$ & -& 7.14 - 7.43 & -           & 9.62 - 10.16  & -            &$0.24\pm0.10$ & -\\
2&J082311.24$-$062408  & $2890\pm800$ & $1880\pm210$& 7.95 - 8.65 & 7.56 - 7.86 & 10.21 - 11.24 & 9.81 - 10.45 &$1.68\pm1.02$ &$0.44\pm0.18$ \\
3&J130817.00$-$344754  & $1640\pm80$ & $1030\pm110$& 7.81 - 8.07 & 6.93 - 7.32 & 10.09 - 10.68 & 9.22 - 9.93  &$1.88\pm0.66$ & $0.29\pm0.16$\\
5&J140050.13$-$291924  & $2130\pm30$ & $1200\pm260$ & 8.35 - 8.53 & 7.12 - 7.46 & 10.64 - 11.14 & 9.40 - 10.06 &$2.60\pm0.61$ & $0.18\pm0.09$\\
6&J141243.15$-$202011  & $2180\pm880$ &-& 8.65 - 9.67 & -           & 9.62 - 10.82  & -            &$0.17\pm0.15$ & -\\
7&J143419.59$-$023543  & $2220\pm340$ &-& 7.38 - 8.25 & -           & 9.95 - 10.68  &-             &$0.54\pm0.29$ & -\\
8&J143931.76$-$372523  & $2680\pm280$ & -& 7.75 - 8.13 &-            & 9.74 - 11.00  & -            &$3.54\pm2.76$ & -\\
9&J150048.73$-$064939  & $2350\pm90$ &$2230\pm150$ & 7.30 - 8.29 & 6.65 - 7.45 & 10.63 - 11.14 & 8.99 - 10.09 &$1.89\pm0.50$ &$0.08\pm0.12$ \\
10&J151003.71$-$220311 & $1460\pm220$ &- & 8.29 - 8.50 & -           & 9.53 - 10.29  & -            &$0.53\pm0.43$ & -\\
11&J151310.42$-$221004 & $3690\pm390$ &-& 7.01 - 7.53 & -           & 10.62 - 11.53 & -            &$0.91\pm0.45$ & -\\
13&J154141.64$-$114409 & $2430\pm470$ &- & 8.52 - 8.87 & -           & 10.24 - 11.08 & -            &$1.99\pm1.68$ & -\\
14&J163426.87$-$172139 & $2680\pm260$ &- & 7.92 - 8.46 & -           & 10.31 - 11.08 & -            &$1.84\pm0.94$ & -\\
16&J195141.22$-$042024 & $1430\pm390$ & $1060\pm120$ & 8.16 - 8.52 & 7.03 - 7.51 & 9.91 - 10.88  & 9.34 - 10.13 & $0.32\pm0.37$& $0.07\pm 0.05$\\
18&J204049.51$-$390400 & - & $490\pm70$ & -            &6.45 - 6.82  &-              & -            & -            &$0.03\pm0.02$ \\
\hline
\end{tabular}
\end{table*}

Although difficult to compare directly due to the different methods used, BH masses presented here in Table \ref{tab:mass} are compared with those from \citet{Lonsdale2015} and \citet{Kim2013} in Fig. \ref{fig:bhcomp}. It shows the range of calculated values and the errors for each of the 18 sources calculated using the \ofive{} lines. We also include masses from our two Balmer line measurements. Of the masses calculated by \citeauthor{Lonsdale2015}, most are in agreement with our \ofive{} calculated masses, with a mean difference of $\sim 0.40$ dex, corresponding to a value of $\rm{\lambda}_{Edd} \sim 0.80$ rather than our value of $\rm{\lambda}_{Edd} = 1$, although their method, based on SED fitting, also varied from ours. Our calculated \ofive{} masses are also mostly in agreement with \citeauthor{Kim2013}, who also used the \ofive{} emission line, with a mean difference of  $\sim 0.46$ dex across a small comparison range. In the case of the Balmer lines, we calculated masses significantly lower than \citeauthor{Lonsdale2015}'s, with a mean difference of $\sim 0.94$ dex for our \halpha{} masses, and \hbeta{} masses a further $\sim 0.90$ dex lower.

Most noticeable is the difference between the \hbeta{} mass estimates compared to all others. Agreeing within error for no sources, and $\sim 1$ order of magnitude lower than the \halpha{} estimates, a modification of the scaling factor in Equation \ref{eq:mhb} by 7.94 would be required to bring the \hbeta{} mass estimates in line with the \halpha{} estimates. These masses have not been corrected for extinction and correcting for this (see Section \ref{sec:extinc}) should make them more consistent. However, due to the lack of detections and high levels of noise in the measured \hbeta{} emission lines, we will assume that the \hbeta{} inferred masses are less reliable and use values consistent with \halpha{}, but will return to discuss the \hbeta{} lines strengths in Section \ref{sec:extinc} in the context of extinction. All masses are yet to be corrected for extinction, and high values of $A_{V}$ will increase all the mass estimates and uncertainties. While there is a large scatter in masses for the majority of these sources, the removal of the \hbeta{} results as less reliable much reduces this scatter and increases the overall confidence of our masses compared with others.

\section{Factors that affect our mass measurements}
\label{sec:discuss}
\subsection{Eddington Ratio}
\label{sec:edd}
The choice of $\rm{\lambda}_{Edd}$ is a major factor affecting our calculated BH masses. A choice of $\rm{\lambda}_{Edd} =1$ represents a classical lower limit to mass if the sources remain in hydrostatic equilibrium; however; it is highly likely that these sources are accreting at unsustainable super-Eddington rates, as suggested in similarly luminous and obscured Hot DOGS \citep{Assef2015,Tsai2015,Tsai2018,Wu2018,Tsai2018}.

By comparing the widths of our measured emission lines presented in Table \ref{tab:spec}, in some cases we find that the \ofive{} lines are comparably broad, if not broader, than the Balmer line widths. This suggests the possibility of outflows broadening some of our measured line widths which may require super-Eddington accretion. Evidence for outflows has been found for Hot DOGS \citep{hotdog2020}, as well as being previously suggested for RWGS \citep{Kim2013}. However, for those sources with comparably broad \ofive{} line widths, it is also possible that the Balmer line emission is actually dominated by the narrow line with the broad line emission being completely obscured. If this was the case then the Balmer line based black hole mass values (and subsequent Eddington rates) will be inaccurate due to the increased distance to the narrow-line region, and the \ofive{} mass values should be used.

By inserting the \halpha{}-derived BH masses into Equation \ref{eq:omass}, we can estimate the Eddington ratio of the sources. We estimate Eddington ratios for the 11 sources with both \halpha{} and \ofive{} line detections and find $\rm{\lambda}_{Edd} = 0.34 - 34.59$ with a mean $\rm{\lambda}_{Edd} = 10.12$. It is therefore plausible that most of our sources are accreting at super-Eddington rates and that \citet{Lonsdale2015} are likely to have over-estimated their BH mass measurements based on their use of $\rm{\lambda}_{Edd} = 0.25$, though their method differs from ours, as they derive masses based on SED fitting. We find one source (W1500) with $\rm{\lambda}_{Edd} \leq 1$, while the remaining 10 sources have $\rm{\lambda}_{Edd} \geq 1$. Within this we measure $1 \leq \rm{\lambda}_{Edd} \leq 5$ for six sources. Our two largest Eddington ratios have $\rm{\lambda}_{Edd} \geq 25$, but whilst these two measurements are greater than the others, they are consistent within errors of each other.

Table \ref{tab:lamedd} lists all of the calculated $\rm{\lambda}_{Edd}$ values and their associated errors. Whilst the $\rm{\lambda}_{Edd}$ values provided here imply that RWGs likely accrete at super-Eddington rates with 10/11 sources having $\rm{\lambda}_{Edd}\geq1$, these values are not all statistically significant. The larger uncertainties on our $\rm{\lambda}_{Edd}$ values mean that the Eddington ratio of $\rm{\lambda}_{Edd} = 1$ remains within $2\sigma$ for two of these 10 sources and within $3\sigma$ for five. Therefore whilst our data does suggests that some RWGs accrete at super-Eddington rates, we can only significantly show this for half of our sources.

\begin{table}
\caption{\label{tab:lamedd} The measured Eddington ratios for the 12 sources with both \halpha{} and \ofive{} line detections.}
\small
\centering
\begin{tabular}{l l l}
\hline
\hline
Source ID &WISE designation & $\rm{\lambda}_{Edd}$ \\
\hline
1& J081131.61$-$222522&  $12.98 \pm 4.83$\\
2& J082311.24$-$062408& $2.07 \pm 0.32$\\
3& J130817.00$-$344754& $25.24 \pm 10.39$\\
5& J140050.13$-$291924& $5.41 \pm 3.24$\\
6& J141243.15$-$202011& $2.09 \pm 0.27$\\
7& J143419.59$-$023543& $10.66 \pm 3.02$\\
9& J150048.73$-$064939& $0.34 \pm 0.18$\\
10& J151003.71$-$220311& $12.06 \pm 2.61$\\
13& J154141.64$-$114409& $3.74 \pm 0.74$\\
14& J163426.87$-$172139& $2.02 \pm 0.61$\\
16& J195141.22$-$042024& $34.59 \pm 5.51$\\
\hline
\end{tabular}
\end{table}

Correcting our masses for these new values of $\rm{\lambda_{Edd}}$ from Table \ref{tab:lamedd} reduce our masses by a mean of $\sim 0.68$ dex, consistent with the \halpha{} masses the $\rm{\lambda_{Edd}}$ values were derived from. Using the \citet{Lonsdale2015} calculated masses to instead estimate Eddington ratio results gives a mean $\rm{\lambda_{Edd}} = 0.95$, with five sources having values of $\rm{\lambda_{Edd}} \geq 1$. Even though the \citeauthor{Lonsdale2015} BH masses may be overestimates of the true value, this demonstrates that once the \halpha{} calculated BH masses are corrected for extinction, and are will be closer to the \citeauthor{Lonsdale2015} values, we will still predict that our sources are accreting at super-Eddington rates.

\begin{figure}
\centering
\includegraphics[width = 0.48\textwidth]{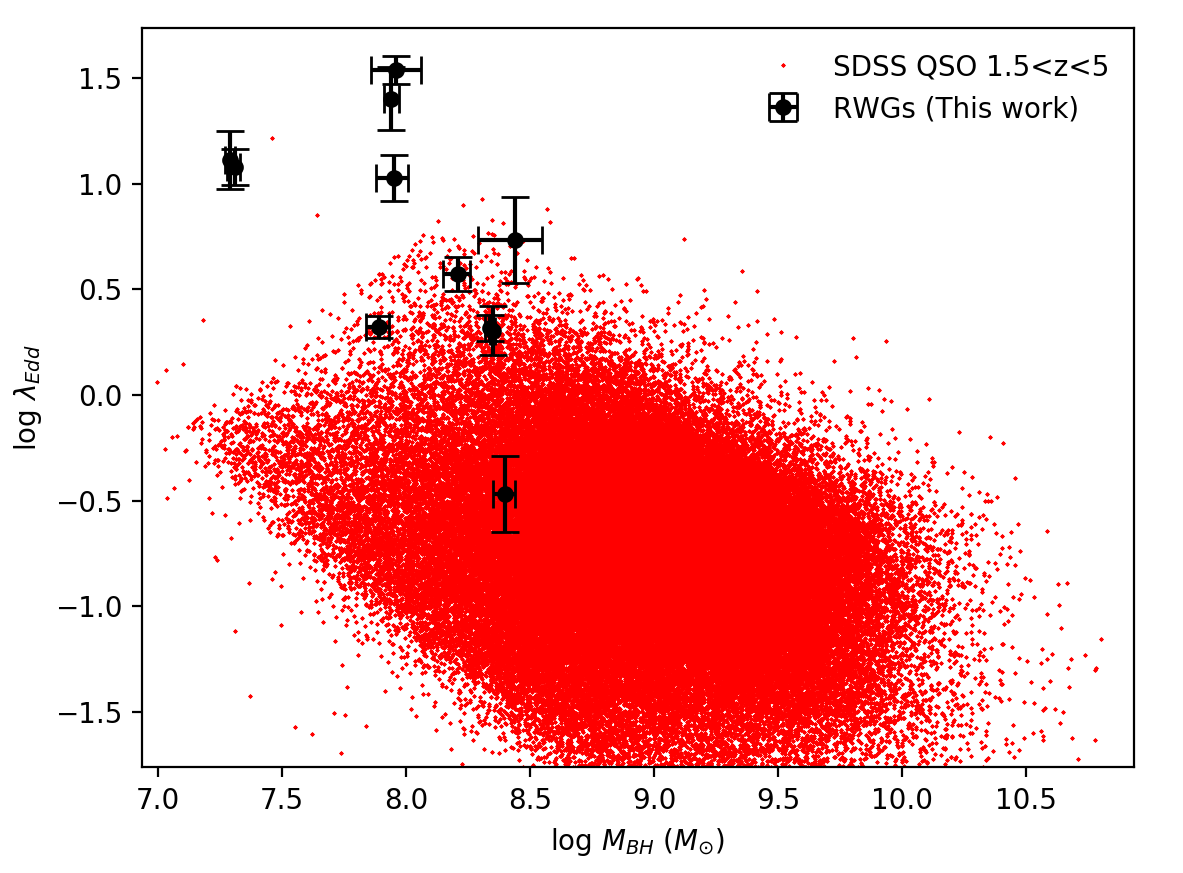}
\caption{\label{fig:bhlamedd} The Eddington ratio needed to match the the \ofive{} emission luminosity, to the BH mass measured from the \halpha{} line. It is plotted against the \halpha{} calculated BH mass. Also included are 20000 typical QSOs from \citet{Shen2011}. Note values plotted are given in Table \ref{tab:lamedd}.}
\end{figure}

Fig. \ref{fig:bhlamedd} shows our measured Eddington ratio, $\rm{\lambda}_{Edd}$, (given in Table \ref{tab:lamedd}) plotted against the \halpha{}-calculated BH mass. Also plotted in red are 20,000 SDSS $1.5 < z < 5$ quasars from \citet{Shen2011} to show how our measurements compare with typical Quasi-Stellar Objects (QSOs), in SDSS DR7. Though our masses are an average 0.8 dex smaller, our sources lie above the mass region of the typical SDSS QSOs, with a mean Eddington ratio $\sim60$ times higher, laying on the upper left of the distribution of Eddington ratios. Five sources with the largest $\rm{\lambda}_{Edd}$ measurements lie completely outside the SDSS population showing that RWGs in general accrete at higher than average Eddington ratios, when compared to typical quasars.

The large uncertainties in our calculated $\rm{\lambda}_{Edd}$ have a variety of origins: the \halpha{}-calculated BH mass and uncertainties in the measurements of the \halpha{} flux and FWHM and the conversion factor from \ofive{} luminosity to bolometric \citep{Shen2011}. The broad \halpha{} component cannot be distinguished from the \nii{} doublet so we assume all the flux comes from the \halpha{}. However, this may not actually be the case and correcting for this might give a different value. Line blending may also have affected the width of the broad line measured and used in this method. These uncertainties contribute to the larger errors on our estimated Eddington ratios; however, in comparison with typical QSOs at similar redshifts \citep{Shen2011}, the RWGs are shown to be accreting at higher than typical Eddington rates. Although our sample is small, we measure the mean of our sample, 10.12, to be $\sim 60$ times bigger than the value of 0.17 from typical QSOs \citep{Shen2011}.

\subsection{Extinction}
\label{sec:extinc}
RWGs are selected to be extremely luminous in the mid-IR, and thus dominated by hot dust emission, likely indicating the proximity of substantial extinction to the central engine. Due to the expected high levels of dust scattering, the calculated optical luminosities above are likely to be underestimates of the true unobscured values. We can test this by measuring the extinction of nuclear emission lines and its effect on our BH and host galaxy masses. As these galaxies are selected to be highly obscured systems these are important corrections that need to be made.

\subsubsection{Extinction estimates from the Balmer decrement}
Following previous work \citep[eg.][]{Dom2013,Kim2013}, the Balmer lines can be used to estimate dust extinction. Any deviation of the \halpha{}:\hbeta{} ratio from the expected optical depth value of 2.86, as expected from Case B recombination, can be used to estimate extinctions. By assuming an extinction law from \citet{Cardelli1989} and a value of $R_{V}=3.1$ \citep{Eisenhardt2012} we can use the Balmer decrement to estimate the $V$-band extinction. In our sample we have 13 detections of the \halpha{} line and six detections of the \hbeta{} line, with all but the W2040 \hbeta{} detection having a companion \halpha{} measurement. We therefore have direct extinction estimates for five of our sources (see Table \ref{tab:mesextinc}).

We measure $A_{V} > 2$ for four of our five sources based on our measured Balmer lines. The extinction measurement of $A_{V} < 0$ for W0823 is however unexpected, as there appears to be a well defined \hbeta{} line implying that it has not been obscured. Any corrections of the \halpha{} spectra due the \nii{} lines would also lead to these negative extinction measurements, but a value of $A_{V} > 0$ is still within errors for this source. 

As shown in Fig. \ref{fig:bhcomp}, in the absence of extinction, BH masses calculated from the \hbeta{} emission lines are considerably lower than those calculated from other emission lines. Correcting these masses for extinction will increase them, narrowing the gap between them the other BH mass measurements. However, before extinction correction there is no agreement within these errors all six of these sources and the BH masses calculated are an average 0.90 dex lower than the next lowest estimate, implying that the \hbeta{} measurements are less reliable, or something else has not been accounted for.  If we now choose to discount our \hbeta{} emission line measurements, we can make tighter constraints on our BH and host masses, and make additional extinction estimates for sources without apparent \hbeta{} lines.

\subsubsection{Extinction estimates from a simulated \hbeta{} line.}
The errors propagated for our extinction values are dominated by the large uncertainties in the noisy and hard to measure \hbeta{} emission lines and this has lead to large uncertainties in our calculated $A_{V}$ and $E(B-V)$ measurements. We have therefore instead used a simulated typical \hbeta{} emission line to attempt to better understand how obscured these sources actually are. With the assumption that all line measurements should give the same BH masses, the \halpha{} BH mass is used to predict an lower limit to flux of a typical \hbeta{} line of fixed width of FWHM = $\rm{1480\,kms^{-1}}$ \footnote{ A FWHM of $\rm{1480\,kms^{-1}}$ was chosen as it is the mean FWHM of our broad \hbeta{} lines.}, through use of reversing Equation \ref{eq:mhb}. The \halpha{} mass was used here instead of the \ofive{} mass due to the \ofive{}'s heavy dependence on the poorly constrained $\rm{\lambda}_{Edd}$ heavily affecting the calculated mass value. This calculated flux lower limit and width are then used to simulate a Gaussian like profile, with noise added to recreate a realistic spectrum. This allows us to then fit the simulated emission line as done for the real data. These measured fluxes and their corresponding errors are then approximately typical of true emission lines and likely lower limits. Whilst not exact, these simulated detections should provide suitable extinction measurements to allow us to estimate a lower limit to how obscured our sources are, and to approximately correct our mass measurements. This simulation was done for the 13 sources with a detected \halpha{} emission line, see Table \ref{tab:simextinc}, to provide 13 simulated \hbeta{} detections for the extinction measurements.

All but one source has positive visual extinction measured, but, for W1513 the negative extinction derived has high uncertainties, due to greater noise in the spectrum, with a positive $A_{V}$ value within error. We also now predict reddening for W0823 with negative extinction previously measured. These calculated extinctions are likely to be representative of the true values agreeing within error to the $V$ band extinction calculated from the measured emission line ratios for W1400. Our errors on these calculated values are larger, owing to the simulation steps, with an average predicted visual extinction for the sources in Table \ref{tab:simextinc} of $A_{V} = 3.62 \pm 2.78$ mag.

Whilst the \halpha{} derived BH mass is used to help scale the simulated \hbeta{} emission line in order to estimate the extinction, it is dependent on the source redshift and FWHM as well as the flux. The simulated \hbeta{} flux that is measured also depends heavily on the assumed FWHM. However, in comparison to the $A_{V}$ values needed to match the BH masses derived by \citet{Lonsdale2015}, 11/13 of our  $A_{V}$ values agree within error. This allows confidence that our simulated emission lines are providing useful extinction estimates.

\subsubsection{Extinction corrected masses}
Due to the wavelength-dependent nature of extinction, it is more difficult to predict the effect this extinction will have on our results. We can, however, briefly explore how large an effect extinction corrections will have on our measured BH masses on a case-by-case basis.

We will illustrate the effect of extinction using W1400 as it has a clear and well measured \ofive{} emission line (see Fig. \ref{fig:spec}), and both Balmer lines measured. The \ofive and \citet{Lonsdale2015} BH mass measurements agree within errors to: $\log \rm{M_{BH}}$ = 9.02 - 9.29 \msun{}, with the \ofive{} line suggesting a mass of $\log \rm{M_{BH} \approx 9.17}$ \msun{} (see Tables \ref{tab:mass} and \ref{tab:appmass}). We now compare this mass to the masses corrected for the extinction we measure using the Balmer lines, the extinction we predict based on a simulated \hbeta\, FWHM of 1480\,$\rm{kms^{-1}}$, and the mean predicted extinction for our 13 sources. Due to the wavelength dependence, the extinction is applied to the original 1D spectra, with the multi-Gaussian models then refitted to the binned spectra\footnote{Extinction is applied to the flux density at each wavelength point of the original 1D spectra which is then binned as before. The original multi-Gaussian models are then refitted to the spectra, providing the extinction corrected flux and FWHM of the emission line.}. Using the \citet{Cardelli1989} extinction law, we correct for our measured extinction of $A_{V} = 3.93^{+0.24}_{-0.25}$ at each wavelength point, leading to a BH mass $\sim 0.45$ dex greater than our original measurement. Correcting for the extinction predicted, by assuming a broad \hbeta{} line of 1480\,$\rm{kms^{-1}}$, we measure a BH mass $\sim 0.52$ dex larger than the original uncorrected mass, with even greater BH masses if even broader lines are simulated. Correcting for the mean value of our simulated extinctions, $A_{V}$ = 3.62, we measure a mass increase of $\sim 0.39 $ dex, with all of the BH masses, host masses and Eddington ratios corrected for the measured, predicted and average extinction of W1400 given in Table \ref{tab:extincmass}. 

This average mass increase would mean a host mass range of $\log \rm{M}_{Host}$ = 10.95 - 11.61 \msun{}. While these BH and host galaxy mass extinction corrected values are not implausible, and now agree with previously measured values, it is still hard to estimate the full effect of extinction. With a measured Eddington ratio of $\rm{\lambda}_{Edd} = 5.41 \pm 3.24$ for W1400, our $\rm{\lambda}_{Edd}$ choice of one means we overestimated our original \ofive{} BH mass which, when corrected for this value of $\rm{\lambda}_{Edd}$, is reduced by $\sim 0.7$ dex. As most of our sources are potentially accreting at super Eddington rates, this should reduce our BH and host masses predicted values. Correcting the \ofive{} W1400 mass for our measured $A_{V}$ and newly extinction corrected Eddington ratio gives $\log \rm{M}_{BH} \sim 8.89$ \msun{}, now in agreement with the \halpha{} calculated BH mass.

We also applied the extinction using the \citet{calzetti2000} extinction law, as a further test, with no difference to our final results. However, it may be that none of these extinction curves represent our unique population, which would affect our measured values. The extinction could also be greyer than the curves used \citep[e.g. ][]{Roebuck2019}, due to radiative transfer effects.

While no exact measurements have been made previously for the extinction of our full source sample, \citet{Kim2013} measure $\log A_{V} = 1.3^{+0.8}_{-0.6}$ for W1400, from \halpha{} and \hbeta{} measurements, measuring the extinction to be much greater than we do. The extremely luminous and obscured nature of these sources means that correct extinction measurements are needed to understand this unique population of galaxies. While briefly explored here our results require a more detailed analysis. With only five measurements of extinction available from our data (due to only five sources having both \halpha{} and \hbeta{} line detections), and our uncertain \hbeta{} measurements, the simulated emission lines will hopefully provide us with better estimates. Only done here for a typical line of FWHM = 1480\,$\rm{kms^{-1}}$, and applied to W1400, full analysis is needed for our entire sample.

We currently predict that our \halpha{} BH masses will be increased by an average of $\sim 0.5$ orders of magnitude due to extinction with masses in the range of $\log\left(\rm{\frac{{M}_{BH}}{\text{M}_{\odot}}}\right)$ = 8.66 - 9.14, matching other \halpha{} calculated masses of radio quiet QSOs at z = 2 \citep{z=2qso}, and SDSS z = 6 QSOs \citep{z=6qso}. Our masses could be scaled with chosen $\rm{\lambda}_{Edd}$ values, but it is clear more investigation in this area is needed. Whilst we estimate less extinction as compared to other work \citep{Kim2013}, and apply the extinction using the standard methods, our outcome is still uncertain and with open questions including whether the standard methodology is valid in this unique situation. Compared to the range of \halpha{} luminosities $ 10^{40} - 10^{44}\,\rm{erg\,s^{-1}}$ used in \citet{Greene2005}, it is possible that our sample with \halpha{} luminosities of $10^{43} - 10^{44}\,\rm{erg\,s^{-1}}$ before extinction, is just too bright and that their relation does not apply for these highest luminosties. With extinction constrained by the total power of our SEDs and possible time variations, new and more complicated methodology may be needed for accurate extinction estimates.

The results presented in this section provide many more questions to be answered in the future. With the launch of the James Webb Space Telescope (JWST) \citep{JWST}, on the horizon, the availability of high resolution near-IR spectra should allow fast advancements in this field and the ability to accurately calculate extinction measurements for these types of extreme obscured galaxies.

{\renewcommand{\arraystretch}{1.5}
\begin{table*}
\small
\centering
\caption{\label{tab:mesextinc} The calculated reddening and extinction (visual and at the wavelengths of the three observed emission lines) for the five sources with both Balmer lines detected.}
\begin{tabular}{l l l l l l l }
\hline
\hline
Source ID & WISE designation & E(B-V) (mag)& \multicolumn{4}{l}{$\rm{A}_{\lambda}$ (mag)} \\
& & & $\rm{\lambda}$ = V & $\rm{\lambda}$ = \halpha{} & $\rm{\lambda}$ = \hbeta{} & $\rm{\lambda}$ = \ofive{} \\
\hline 
2&J082311.24$-$062408 &$-0.03^{+0.21}_{-0.24}$ &  $-0.11^{+0.64}_{-0.74}$&  $-0.09^{+0.52}_{-0.61}$ &  $-0.12^{+0.74}_{-0.57}$ &  $-0.12^{+0.74}_{-0.57}$\\
3&J130817.00$-$344754 &$0.60^{+0.39}_{-0.35}$  & $1.86^{+1.21}_{-1.09}$  &  $1.52^{+0.99}_{-0.89}$ &  $2.16^{+1.42}_{-1.54}$ &  $2.08^{+1.36}_{-1.46}$\\
5&J140050.13$-$291924 &$1.27^{+0.07}_{-0.08}$  & $3.93^{+0.24}_{-0.25}$  &  $3.21^{+0.20}_{-0.20}$ &  $4.58^{+0.28}_{-0.29}$ &  $4.40^{+0.27}_{-0.28}$\\
9&J150048.73$-$064939 &$1.84^{+0.07}_{-0.07}$  &  $5.70^{+0.23}_{-0.23}$ &  $4.66^{+0.19}_{-0.19}$ &  $6.64^{+0.27}_{-0.28}$ &  $6.39^{+0.26}_{-0.27}$\\
16&J195141.22$-$042024&$0.65^{+0.13}_{-0.14}$ & $2.01^{+0.39}_{-0.43}$ &  $1.64^{+0.32}_{-0.35}$ &  $ 2.34^{+0.45}_{-0.49}$ &  $ 2.26^{+0.44}_{-0.45}$\\
\hline 
\end{tabular}
\end{table*}}

{\renewcommand{\arraystretch}{1.5}
\begin{table*}
\small
\centering
\caption{\label{tab:simextinc} The calculated reddening and extinction (visual and at the wavelengths of the three observed emission lines) for the thirteen sources with an \halpha{} measurement and a simulated (FWHM = 1480 $\rm{kms^{-1}}$) \hbeta{} line.}
\begin{tabular}{l l l l l l l}
\hline
\hline
Source ID & WISE designation & E(B-V) (mag)& \multicolumn{4}{l}{$\rm{A}_{\lambda}$ (mag)} \\
& & & $\rm{\lambda}$ = V & $\rm{\lambda}$ = \halpha{} & $\rm{\lambda}$ = \hbeta{} & $\rm{\lambda}$ = \ofive{} \\
\hline 
1&J081131.61$-$222522  &$ 2.31 ^{+ 0.91 }_{- 0.51 }$ &  $ 7.17 ^{+ 2.82 }_{- 1.59 }$ & $ 5.86 ^{+ 2.31 }_{- 1.30 }$ & $ 8.34 ^{+ 3.29 }_{- 3.79 }$ & $ 8.03 ^{+ 3.16 }_{- 3.47 }$ \\
2&J082311.24$-$062408  &$ 0.24 ^{+ 1.23 }_{- 1.02 }$ &  $ 0.73 ^{+ 3.81 }_{- 3.18 }$ & $ 0.60 ^{+ 2.99 }_{- 2.60 }$ & $ 0.86 ^{+ 4.02 }_{- 2.85 }$ & $ 0.82 ^{+ 3.94 }_{- 2.82 }$  \\
3&J130817.00$-$344754  &$ 2.10 ^{+ 0.68 }_{- 0.50 }$ &  $ 6.52 ^{+ 2.11 }_{- 1.55 }$ & $ 5.33 ^{+ 1.72 }_{- 1.26 }$ & $ 7.60 ^{+ 2.45 }_{- 3.53 }$ & $ 7.31 ^{+ 2.36 }_{- 3.24 }$  \\
5&J140050.13$-$291924  &$ 1.39 ^{+ 0.26 }_{- 0.23 }$ &  $ 4.32 ^{+ 0.82 }_{- 0.71 }$ & $ 3.53 ^{+ 0.67 }_{- 0.58 }$ & $ 5.03 ^{+ 0.96 }_{- 2.08 }$ & $ 4.84 ^{+ 0.92 }_{- 1.89 }$ \\
6&J141243.15$-$202011  &$ 1.07 ^{+ 1.05 }_{- 1.04 }$ &  $ 3.32 ^{+ 3.26 }_{- 3.24 }$ & $ 2.72 ^{+ 2.67 }_{- 2.65 }$ & $ 3.87 ^{+ 3.94 }_{- 3.80 }$ & $ 3.72 ^{+ 3.81 }_{- 3.65 }$  \\
7&J143419.59$-$023543  &$ 1.02 ^{+ 0.76 }_{- 0.52 }$ &  $ 3.15 ^{+ 2.35 }_{- 1.62 }$ & $ 2.58 ^{+ 1.74 }_{- 1.32 }$ & $ 3.67 ^{+ 2.91 }_{- 2.42 }$ & $ 3.53 ^{+ 2.76 }_{- 2.28 }$  \\
8&J143931.76$-$372523  &$ 0.34 ^{+ 0.30 }_{- 0.28 }$ &  $ 1.04 ^{+ 0.93 }_{- 0.87 }$ & $ 0.85 ^{+ 0.30 }_{- 0.26 }$ & $ 1.21 ^{+ 0.64 }_{- 0.62 }$ & $ 1.17 ^{+ 0.59 }_{- 0.57 }$ \\
9&J150048.73$-$064939  &$ 1.00 ^{+ 0.33 }_{- 0.26 }$ &  $ 3.09 ^{+ 1.02 }_{- 0.81 }$ & $ 2.53 ^{+ 0.83 }_{- 0.66 }$ & $ 3.6 ^{+ 1.18 }_{- 1.74 }$ & $ 3.46 ^{+ 1.14 }_{- 1.6 }$   \\
10&J151003.71$-$220311 &$ 2.25 ^{+ 1.90 }_{- 1.50 }$ &  $ 6.99 ^{+ 5.90 }_{- 4.65 }$ & $ 5.71 ^{+ 4.36 }_{- 3.80 }$ & $ 8.14 ^{+ 8.18 }_{- 6.22 }$ & $ 7.83 ^{+ 7.57 }_{- 5.92 }$ \\
11&J151310.42$-$221004 &$ -0.48 ^{+ 0.76 }_{- 0.49 }$ & $ -1.50 ^{+ 2.36}_{- 1.53 }$ & $ -1.22 ^{+ 1.93 }_{- 1.25 }$ & $ -1.74 ^{+ 2.74 }_{- 0.73 }$ & $ -1.68 ^{+ 2.64 }_{- 0.80 }$ \\
13&J154141.64$-$114409 &$ 0.81 ^{+ 0.87 }_{- 0.86 }$ &  $ 2.50 ^{+ 2.70 }_{- 2.66 }$ & $ 2.04 ^{+ 2.77 }_{- 2.17 }$ & $ 2.91 ^{+ 4.06 }_{- 3.04 }$ & $ 2.80 ^{+ 3.65 }_{- 2.93 }$   \\
14&J163426.87$-$172139 &$ 0.51 ^{+ 0.47 }_{- 0.62 }$ &  $ 1.58 ^{+ 1.47 }_{- 0.92 }$ & $ 1.29 ^{+ 0.93 }_{- 0.57 }$ & $ 1.84 ^{+ 1.95 }_{- 1.11 }$ & $ 1.77 ^{+ 1.82 }_{- 1.05 }$ \\
16&J195141.22$-$042024 &$ 2.63 ^{+ 0.99 }_{- 0.95 }$ &  $ 8.14 ^{+ 3.07 }_{- 2.95 }$ & $ 6.66 ^{+ 2.51 }_{- 2.41 }$ & $ 9.48 ^{+ 5.43 }_{- 5.24 }$ & $ 9.12 ^{+ 5.05 }_{- 4.88 }$  \\
\hline
\end{tabular}
\end{table*}}

\begin{table}
\caption{ \label{tab:extincmass} The corrected BH masses, host masses and Eddington ratios of W1400. These are calculated when our spectra are corrected for our measured, predicted and average $\rm{A_{V}}$ values.}
\small
\centering
\begin{tabular}{l l l l}
\hline
\hline
$\rm{A_{V}}$ (mag) & $\log\left(\rm{\frac{{M}_{BH}}{\text{M}_{\odot}}}\right)$ & $\log\left(\rm{\frac{{M}_{Host}}{\text{M}_{\odot}}}\right)$& $\rm{\lambda}_{Edd}$ \\
\hline
3.93 &  8.71 - 9.08 & 10.99 - 11.69 & $110 \pm 60$ \\
4.32 & 8.79 - 9.14  & 11.07 - 11.75 & $140 \pm 80$ \\
3.62 & 8.66 - 9.01 & 10.95 - 11.61  & $90 \pm 50$ \\

\hline
\end{tabular}
\end{table}

\section{Conclusions}
\label{sec:conc}
The results of the VLT X-shooter near-IR spectroscopy and ISAAC imaging on 30 radio-selected, extremely luminous WISE galaxies are as follows: 

i) We present flux densities from $J$ and ${K}_{s}$ band imaging performed with ISAAC for 30 radio selected, luminous IR WISE galaxies, alongside WISE and ALMA flux densities previously obtained.

ii) We detect emission lines in 18/27 of our sources observed with X-shooter. This breaks down into 13 \halpha{} detections, six \hbeta{} and 15 \ofive{} detections. Five of our sources have both \halpha{} and \hbeta{} detections allowing for extinction to be estimated from the Balmer decrement.

iii) We additionally detect 10 \lya{} and five \civ{} lines in the UVB spectra.

iv) We assume $\rm{\lambda}_{Edd} = 1 $ to determine a lower limit on the BH masses of 15 galaxies using the luminosity of the \ofive{} emission line as a proxy for bolometric luminosity. We measure BH masses in the range $\log\left(\rm{\frac{{M}_{BH}}{\text{M}_{\odot}}}\right)$ = 7.9 - 9.4 with corresponding host masses in the range $\log\left(\rm{\frac{{M}_{Host}}{\text{M}_{\odot}}}\right)$ = 10.4 - 12.0.

v) We use the \halpha{} and \hbeta{} emission lines to provide additional, independent measurements of BH and host masses, with \halpha{} determined BH masses of $\log\left(\rm{\frac{{M}_{BH}}{\text{M}_{\odot}}}\right)$ = 7.3 - 8.7.

vi) We use the \ofive{} line luminosities and the \halpha{} calculated BH masses to predict that most RWGs may accrete at super Eddington rates, with a mean value of $\rm{\lambda}_{Edd} = 10.12 $ from 11 of our sources ranging from $\rm{\lambda}_{Edd} = 0.3 - 34.6$. However a value of $\rm{\lambda}_{Edd} \geq 1$ is only statistically significant by $3 \sigma$ for five sources.

vii) We measure a mean visual extinction of $A_{V}$ = 2.68 mag for five of our sources using the Balmer decrement. We then infer an average extinction of $A_{V}$ = 3.62 mag and reddening, E(B-V) = 1.39 mag for all sources, by simulating a typical emission line. These corrections increase our measured masses by $\sim$ 0.5 orders of magnitude in the most obscured cases.

\section*{Acknowledgements}

We thank the referee for their useful comments allowing for improvements to this manuscript. The authors also wish to thank the staff at the Department of Physics and Astronomy, University of Leicester for their support. ERF is supported by a Science and Technologies Facilities Council (STFC) studentship. RJA was supported by FONDECYT grant number 1191124. MK was supported by the National Research Foundation of Korea (NRF) grant funded by the Korea government (MSIT) (No. 2020R1A2C4001753). This work is based on observations collected at the European Southern Observatory under ESO programmes 091.A-0545(A) and 290.A-5042(A), and the authors also thank those involved with the original ESO proposals.


\section*{Data Availability}

The data underlying this article was collected at the European Southern Observatory under ESO programmes 091.A-0545(A) and 290.A-5042(A) and will be shared on reasonable request to the corresponding author.



\bibliographystyle{mnras}
\bibliography{bib.bib} 

\begin{thebibliography}{}
\makeatletter
\relax
\def\mn@urlcharsother{\let\do\@makeother \do\$\do\&\do\#\do\^\do\_\do\%\do\~}
\def\mn@doi{\begingroup\mn@urlcharsother \@ifnextchar [ {\mn@doi@}
  {\mn@doi@[]}}
\def\mn@doi@[#1]#2{\def\@tempa{#1}\ifx\@tempa\@empty \href
  {http://dx.doi.org/#2} {doi:#2}\else \href {http://dx.doi.org/#2} {#1}\fi
  \endgroup}
\def\mn@eprint#1#2{\mn@eprint@#1:#2::\@nil}
\def\mn@eprint@arXiv#1{\href {http://arxiv.org/abs/#1} {{\tt arXiv:#1}}}
\def\mn@eprint@dblp#1{\href {http://dblp.uni-trier.de/rec/bibtex/#1.xml}
  {dblp:#1}}
\def\mn@eprint@#1:#2:#3:#4\@nil{\def\@tempa {#1}\def\@tempb {#2}\def\@tempc
  {#3}\ifx \@tempc \@empty \let \@tempc \@tempb \let \@tempb \@tempa \fi \ifx
  \@tempb \@empty \def\@tempb {arXiv}\fi \@ifundefined
  {mn@eprint@\@tempb}{\@tempb:\@tempc}{\expandafter \expandafter \csname
  mn@eprint@\@tempb\endcsname \expandafter{\@tempc}}}

\bibitem[\protect\citeauthoryear{{Assef} et~al.,}{{Assef}
  et~al.}{2011}]{Assef2011}
{Assef} R.~J.,  et~al., 2011, \mn@doi [\apj] {10.1088/0004-637X/728/1/56},
  \href {http://adsabs.harvard.edu/abs/2011ApJ...728...56A} {728, 56}

\bibitem[\protect\citeauthoryear{{Assef} et~al.,}{{Assef}
  et~al.}{2015}]{Assef2015}
{Assef} R.~J.,  et~al., 2015, \mn@doi [\apj] {10.1088/0004-637X/804/1/27},
  \href {http://adsabs.harvard.edu/abs/2015ApJ...804...27A} {804, 27}

\bibitem[\protect\citeauthoryear{{Becker}, {White}  \& {Helfand}}{{Becker}
  et~al.}{1995}]{Becker1995}
{Becker} R.~H.,  {White} R.~L.,   {Helfand} D.~J.,  1995, \mn@doi [\apj]
  {10.1086/176166}, \href {http://adsabs.harvard.edu/abs/1995ApJ...450..559B}
  {450, 559}

\bibitem[\protect\citeauthoryear{{Bentz} et~al.,}{{Bentz}
  et~al.}{2009}]{bentz2009}
{Bentz} M.~C.,  et~al., 2009, \mn@doi [\apj] {10.1088/0004-637X/705/1/199},
  \href {https://ui.adsabs.harvard.edu/abs/2009ApJ...705..199B} {705, 199}

\bibitem[\protect\citeauthoryear{{Bertin} \& {Arnouts}}{{Bertin} \&
  {Arnouts}}{1996}]{Bertin1996}
{Bertin} E.,  {Arnouts} S.,  1996, \mn@doi [\aaps] {10.1051/aas:1996164}, \href
  {http://adsabs.harvard.edu/abs/1996A26AS..117..393B} {117, 393}

\bibitem[\protect\citeauthoryear{Blain, Smail, Ivison, Kneib  \& Frayer}{Blain
  et~al.}{2002}]{Blain2002}
Blain A.~W.,  Smail I.,  Ivison R.,  Kneib J.-P.,   Frayer D.~T.,  2002,
  \mn@doi [Physics Reports] {https://doi.org/10.1016/S0370-1573(02)00134-5},
  369, 111

\bibitem[\protect\citeauthoryear{{Calzetti}, {Armus}, {Bohlin}, {Kinney},
  {Koornneef}  \& {Storchi-Bergmann}}{{Calzetti} et~al.}{2000}]{calzetti2000}
{Calzetti} D.,  {Armus} L.,  {Bohlin} R.~C.,  {Kinney} A.~L.,  {Koornneef} J.,
   {Storchi-Bergmann} T.,  2000, \mn@doi [\apj] {10.1086/308692}, \href
  {http://adsabs.harvard.edu/abs/2000ApJ...533..682C} {533, 682}

\bibitem[\protect\citeauthoryear{{Cardelli}, {Clayton}  \& {Mathis}}{{Cardelli}
  et~al.}{1989}]{Cardelli1989}
{Cardelli} J.~A.,  {Clayton} G.~C.,   {Mathis} J.~S.,  1989, \mn@doi [\apj]
  {10.1086/167900}, \href {http://adsabs.harvard.edu/abs/1989ApJ...345..245C}
  {345, 245}

\bibitem[\protect\citeauthoryear{{Casey}, {Narayanan}  \& {Cooray}}{{Casey}
  et~al.}{2014}]{Casey2014}
{Casey} C.~M.,  {Narayanan} D.,   {Cooray} A.,  2014, \mn@doi [\physrep]
  {10.1016/j.physrep.2014.02.009}, \href
  {http://adsabs.harvard.edu/abs/2014PhR...541...45C} {541, 45}

\bibitem[\protect\citeauthoryear{{Chiang}, {Overzier}, {Gebhardt}  \&
  {Henriques}}{{Chiang} et~al.}{2017}]{Chiang17}
{Chiang} Y.-K.,  {Overzier} R.~A.,  {Gebhardt} K.,   {Henriques} B.,  2017,
  \mn@doi [\apj] {10.3847/2041-8213/aa7e7b}, \href
  {https://ui.adsabs.harvard.edu/abs/2017ApJ...844L..23C} {844, L23}

\bibitem[\protect\citeauthoryear{{Condon}, {Cotton}, {Greisen}, {Yin},
  {Perley}, {Taylor}  \& {Broderick}}{{Condon} et~al.}{1998}]{Condon1998}
{Condon} J.~J.,  {Cotton} W.~D.,  {Greisen} E.~W.,  {Yin} Q.~F.,  {Perley}
  R.~A.,  {Taylor} G.~B.,   {Broderick} J.~J.,  1998, \mn@doi [\aj]
  {10.1086/300337}, \href {http://adsabs.harvard.edu/abs/1998AJ....115.1693C}
  {115, 1693}

\bibitem[\protect\citeauthoryear{{D{\'{\i}}az-Santos}
  et~al.,}{{D{\'{\i}}az-Santos} et~al.}{2016}]{Diaz-santos2016}
{D{\'{\i}}az-Santos} T.,  et~al., 2016, \mn@doi [\apjl]
  {10.3847/2041-8205/816/1/L6}, \href
  {http://adsabs.harvard.edu/abs/2016ApJ...816L...6D} {816, L6}

\bibitem[\protect\citeauthoryear{{D{\'{\i}}az-Santos}
  et~al.,}{{D{\'{\i}}az-Santos} et~al.}{2018}]{diazsantos18}
{D{\'{\i}}az-Santos} T.,  et~al., 2018, \mn@doi [Science]
  {10.1126/science.aap7605}, \href
  {http://adsabs.harvard.edu/abs/2018Sci...362.1034D} {362, 1034}

\bibitem[\protect\citeauthoryear{{Dom{\'{\i}}nguez} et~al.,}{{Dom{\'{\i}}nguez}
  et~al.}{2013}]{Dom2013}
{Dom{\'{\i}}nguez} A.,  et~al., 2013, \mn@doi [\apj]
  {10.1088/0004-637X/763/2/145}, \href
  {http://adsabs.harvard.edu/abs/2013ApJ...763..145D} {763, 145}

\bibitem[\protect\citeauthoryear{{Eisenhardt} et~al.,}{{Eisenhardt}
  et~al.}{2012}]{Eisenhardt2012}
{Eisenhardt} P.~R.~M.,  et~al., 2012, \mn@doi [\apj]
  {10.1088/0004-637X/755/2/173}, \href
  {http://adsabs.harvard.edu/abs/2012ApJ...755..173E} {755, 173}

\bibitem[\protect\citeauthoryear{{Ferrarese} \& {Merritt}}{{Ferrarese} \&
  {Merritt}}{2000}]{Merrit2000}
{Ferrarese} L.,  {Merritt} D.,  2000, \mn@doi [\apjl] {10.1086/312838}, \href
  {https://ui.adsabs.harvard.edu/abs/2000ApJ...539L...9F} {539, L9}

\bibitem[\protect\citeauthoryear{Gardner et~al.,}{Gardner et~al.}{2006}]{JWST}
Gardner J.~P.,  et~al., 2006, \mn@doi [Space Science Reviews]
  {10.1007/s11214-006-8315-7}, 123, 485

\bibitem[\protect\citeauthoryear{{Gebhardt} et~al.,}{{Gebhardt}
  et~al.}{2000}]{Gebhardt2000}
{Gebhardt} K.,  et~al., 2000, \mn@doi [\apjl] {10.1086/312840}, \href
  {https://ui.adsabs.harvard.edu/abs/2000ApJ...539L..13G} {539, L13}

\bibitem[\protect\citeauthoryear{{Greene} \& {Ho}}{{Greene} \&
  {Ho}}{2005}]{Greene2005}
{Greene} J.~E.,  {Ho} L.~C.,  2005, \mn@doi [\apj] {10.1086/431897}, \href
  {http://adsabs.harvard.edu/abs/2005ApJ...630..122G} {630, 122}

\bibitem[\protect\citeauthoryear{Hatch et~al.,}{Hatch et~al.}{2014}]{Hatch14}
Hatch N.~A.,  et~al., 2014, \mn@doi [\mnras] {10.1093/mnras/stu1725}, 445, 280

\bibitem[\protect\citeauthoryear{Hopkins, Hernquist, Cox, Matteo, Robertson  \&
  Springel}{Hopkins et~al.}{2006}]{Hopkins2006}
Hopkins P.~F.,  Hernquist L.,  Cox T.~J.,  Matteo T.~D.,  Robertson B.,
  Springel V.,  2006, \mn@doi [The Astrophysical Journal Supplement Series]
  {10.1086/499298}, 163, 1

\bibitem[\protect\citeauthoryear{{Jones} et~al.,}{{Jones}
  et~al.}{2015}]{Jones2015}
{Jones} S.~F.,  et~al., 2015, \mn@doi [\mnras] {10.1093/mnras/stv214}, \href
  {http://adsabs.harvard.edu/abs/2015MNRAS.448.3325J} {448, 3325}

\bibitem[\protect\citeauthoryear{{Jun}, {Im}, {Kim}  \& {Stern}}{{Jun}
  et~al.}{2017}]{Jun17}
{Jun} H.~D.,  {Im} M.,  {Kim} D.,   {Stern} D.,  2017, \mn@doi [\apj]
  {10.3847/1538-4357/aa63f9}, \href
  {http://adsabs.harvard.edu/abs/2017ApJ...838...41J} {838, 41}

\bibitem[\protect\citeauthoryear{{Jun} et~al.,}{{Jun}
  et~al.}{2020}]{hotdog2020}
{Jun} H.~D.,  et~al., 2020, \mn@doi [\apj] {10.3847/1538-4357/ab5e7b}, \href
  {https://ui.adsabs.harvard.edu/abs/2020ApJ...888..110J} {888, 110}

\bibitem[\protect\citeauthoryear{{Kaspi}, {Smith}, {Netzer}, {Maoz}, {Jannuzi}
  \& {Giveon}}{{Kaspi} et~al.}{2000}]{Kaspi2000}
{Kaspi} S.,  {Smith} P.~S.,  {Netzer} H.,  {Maoz} D.,  {Jannuzi} B.~T.,
  {Giveon} U.,  2000, \mn@doi [\apj] {10.1086/308704}, \href
  {http://adsabs.harvard.edu/abs/2000ApJ...533..631K} {533, 631}

\bibitem[\protect\citeauthoryear{{Kim}, {Ho}, {Lonsdale}, {Lacy}, {Blain}  \&
  {Kimball}}{{Kim} et~al.}{2013}]{Kim2013}
{Kim} M.,  {Ho} L.~C.,  {Lonsdale} C.~J.,  {Lacy} M.,  {Blain} A.~W.,
  {Kimball} A.~E.,  2013, \mn@doi [\apjl] {10.1088/2041-8205/768/1/L9}, \href
  {http://adsabs.harvard.edu/abs/2013ApJ...768L...9K} {768, L9}

\bibitem[\protect\citeauthoryear{{Kormendy} \& {Ho}}{{Kormendy} \&
  {Ho}}{2013}]{Kormendy13}
{Kormendy} J.,  {Ho} L.~C.,  2013, \mn@doi [\araa]
  {10.1146/annurev-astro-082708-101811}, \href
  {http://adsabs.harvard.edu/abs/2013ARA26A..51..511K} {51, 511}

\bibitem[\protect\citeauthoryear{Kurk et~al.,}{Kurk et~al.}{2008}]{z=6qso}
Kurk J.,  et~al., 2008, \mn@doi [The Astrophysical Journal] {10.1086/521596},
  669, 32

\bibitem[\protect\citeauthoryear{Labita, Decarli, Treves, Falomo, Kotilainen
  \& Scarpa}{Labita et~al.}{2010}]{Decarli10}
Labita M.,  Decarli R.,  Treves A.,  Falomo R.,  Kotilainen J.~K.,   Scarpa R.,
   2010, \mn@doi [\mnras] {10.1111/j.1365-2966.2009.16049.x}, 402, 2453

\bibitem[\protect\citeauthoryear{Laor \& Stern}{Laor \& Stern}{2012}]{stern12}
Laor A.,  Stern J.,  2012, \mn@doi [\mnras] {10.1111/j.1365-2966.2012.21772.x},
  426, 2703

\bibitem[\protect\citeauthoryear{{Lonsdale} et~al.,}{{Lonsdale}
  et~al.}{2015}]{Lonsdale2015}
{Lonsdale} C.~J.,  et~al., 2015, \mn@doi [\apj] {10.1088/0004-637X/813/1/45},
  \href {http://adsabs.harvard.edu/abs/2015ApJ...813...45L} {813, 45}

\bibitem[\protect\citeauthoryear{Lonsdale et~al.,}{Lonsdale
  et~al.}{2016}]{lonsdale16}
Lonsdale C.,  et~al., 2016, \mn@doi [Astronomische Nachrichten]
  {10.1002/asna.201512283}, 337, 194

\bibitem[\protect\citeauthoryear{{Magorrian} et~al.,}{{Magorrian}
  et~al.}{1998}]{Magorrian1998}
{Magorrian} J.,  et~al., 1998, \mn@doi [\aj] {10.1086/300353}, \href
  {https://ui.adsabs.harvard.edu/abs/1998AJ....115.2285M} {115, 2285}

\bibitem[\protect\citeauthoryear{{Moorwood} et~al.,}{{Moorwood}
  et~al.}{1998}]{Moowood1998}
{Moorwood} A.,  et~al., 1998, The Messenger, \href
  {http://adsabs.harvard.edu/abs/1998Msngr..94....7M} {94, 7}

\bibitem[\protect\citeauthoryear{{Orellana, G.}, {Nagar, N. M.}, {Isaak, K.
  G.}, {Priddey, R.}, {Maiolino, R.}, {McMahon, R.}, {Marconi, A.}  \& {Oliva,
  E.}}{{Orellana, G.} et~al.}{2011}]{z=2qso}
{Orellana, G.} {Nagar, N. M.} {Isaak, K. G.} {Priddey, R.} {Maiolino, R.}
  {McMahon, R.} {Marconi, A.}  {Oliva, E.} 2011, \mn@doi [A\&A]
  {10.1051/0004-6361/201015807}, 531, A128

\bibitem[\protect\citeauthoryear{{Penney} et~al.,}{{Penney}
  et~al.}{2019}]{Jordan}
{Penney} J.~I.,  et~al., 2019, \mn@doi [\mnras] {10.1093/mnras/sty3128}, \href
  {http://adsabs.harvard.edu/abs/2019MNRAS.483..514P} {483, 514}

\bibitem[\protect\citeauthoryear{Roebuck, Sajina, Hayward, Martis, Marchesini,
  Krefting  \& Pope}{Roebuck et~al.}{2019}]{Roebuck2019}
Roebuck E.,  Sajina A.,  Hayward C.~C.,  Martis N.,  Marchesini D.,  Krefting
  N.,   Pope A.,  2019, \mn@doi [The Astrophysical Journal]
  {10.3847/1538-4357/ab2bf5}, 881, 18

\bibitem[\protect\citeauthoryear{{Shen} et~al.,}{{Shen}
  et~al.}{2011}]{Shen2011}
{Shen} Y.,  et~al., 2011, \mn@doi [\apjs] {10.1088/0067-0049/194/2/45}, \href
  {http://adsabs.harvard.edu/abs/2011ApJS..194...45S} {194, 45}

\bibitem[\protect\citeauthoryear{{Silva}, {Sajina}, {Lonsdale}  \&
  {Lacy}}{{Silva} et~al.}{2015}]{Silva2015}
{Silva} A.,  {Sajina} A.,  {Lonsdale} C.,   {Lacy} M.,  2015, \mn@doi [\apjl]
  {10.1088/2041-8205/806/2/L25}, \href
  {http://adsabs.harvard.edu/abs/2015ApJ...806L..25S} {806, L25}

\bibitem[\protect\citeauthoryear{{Skrutskie} et~al.,}{{Skrutskie}
  et~al.}{2006}]{2mass}
{Skrutskie} M.~F.,  et~al., 2006, \mn@doi [\aj] {10.1086/498708}, \href
  {http://adsabs.harvard.edu/abs/2006AJ....131.1163S} {131, 1163}

\bibitem[\protect\citeauthoryear{{Tody}}{{Tody}}{1986}]{IRAF}
{Tody} D.,  1986, in {Crawford} D.~L.,  ed.,  \procspie Vol. 627,
  Instrumentation in astronomy VI. p.~733, \mn@doi{10.1117/12.968154}

\bibitem[\protect\citeauthoryear{{Tsai} et~al.,}{{Tsai}
  et~al.}{2015}]{Tsai2015}
{Tsai} C.-W.,  et~al., 2015, \mn@doi [\apj] {10.1088/0004-637X/805/2/90}, \href
  {http://adsabs.harvard.edu/abs/2015ApJ...805...90T} {805, 90}

\bibitem[\protect\citeauthoryear{{Tsai} et~al.,}{{Tsai}
  et~al.}{2018}]{Tsai2018}
{Tsai} C.-W.,  et~al., 2018, \mn@doi [\apj] {10.3847/1538-4357/aae698}, \href
  {http://adsabs.harvard.edu/abs/2018ApJ...868...15T} {868, 15}

\bibitem[\protect\citeauthoryear{{Vernet} et~al.,}{{Vernet}
  et~al.}{2011}]{Vernet2011}
{Vernet} J.,  et~al., 2011, \mn@doi [\aap] {10.1051/0004-6361/201117752}, \href
  {http://adsabs.harvard.edu/abs/2011A%26A...536A.105V} {536, A105}

\bibitem[\protect\citeauthoryear{{Wright} et~al.,}{{Wright}
  et~al.}{2010}]{Wright2010}
{Wright} E.~L.,  et~al., 2010, \mn@doi [\aj] {10.1088/0004-6256/140/6/1868},
  \href {http://adsabs.harvard.edu/abs/2010AJ....140.1868W} {140, 1868}

\bibitem[\protect\citeauthoryear{{Wu} et~al.,}{{Wu} et~al.}{2012}]{Wu2012}
{Wu} J.,  et~al., 2012, \mn@doi [\apj] {10.1088/0004-637X/756/1/96}, \href
  {http://adsabs.harvard.edu/abs/2012ApJ...756...96W} {756, 96}

\bibitem[\protect\citeauthoryear{{Wu} et~al.,}{{Wu} et~al.}{2014}]{Wu2014}
{Wu} J.,  et~al., 2014, \mn@doi [\apj] {10.1088/0004-637X/793/1/8}, \href
  {http://adsabs.harvard.edu/abs/2014ApJ...793....8W} {793, 8}

\bibitem[\protect\citeauthoryear{{Wu} et~al.,}{{Wu} et~al.}{2018}]{Wu2018}
{Wu} J.,  et~al., 2018, \mn@doi [\apj] {10.3847/1538-4357/aa9ff3}, \href
  {http://adsabs.harvard.edu/abs/2018ApJ...852...96W} {852, 96}

\bibitem[\protect\citeauthoryear{{Wylezalek} et~al.,}{{Wylezalek}
  et~al.}{2013}]{Wylezalek2013}
{Wylezalek} D.,  et~al., 2013, \mn@doi [\apj] {10.1088/0004-637X/769/1/79},
  \href {http://adsabs.harvard.edu/abs/2013ApJ...769...79W} {769, 79}

\bibitem[\protect\citeauthoryear{{Zakamska} et~al.,}{{Zakamska}
  et~al.}{2016}]{zakamska16}
{Zakamska} N.~L.,  et~al., 2016, \mn@doi [\mnras] {10.1093/mnras/stw718}, \href
  {https://ui.adsabs.harvard.edu/abs/2016MNRAS.459.3144Z} {459, 3144}

\bibitem[\protect\citeauthoryear{{da Cunha}, {Charlot}  \& {Elbaz}}{{da Cunha}
  et~al.}{2008}]{Dacunha2008}
{da Cunha} E.,  {Charlot} S.,   {Elbaz} D.,  2008, \mn@doi [\mnras]
  {10.1111/j.1365-2966.2008.13535.x}, \href
  {http://adsabs.harvard.edu/abs/2008MNRAS.388.1595D} {388, 1595}

\makeatother
\end{thebibliography}



\appendix

\section{1D spectra}
\label{sec:app}
Given below in Fig. \ref{fig:specap} are the full 1D spectra for all 18 sources with detected emission lines. Note that bins with a flux error $\geq\,3 \times$ their value are masked.

\begin{figure*}
\centering
\includegraphics[width = 0.9\textwidth]{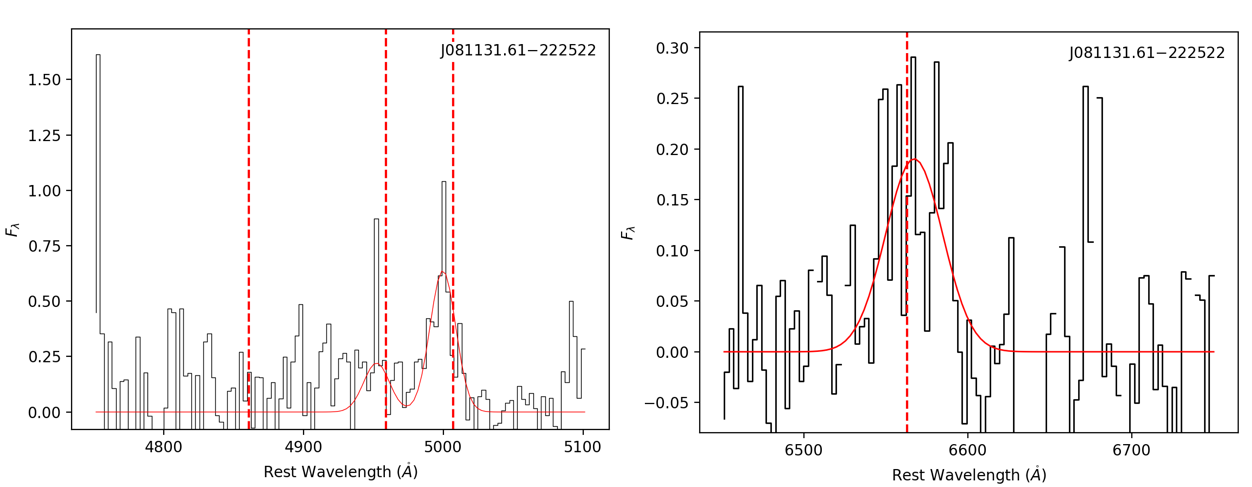}
\includegraphics[width = 0.9\textwidth]{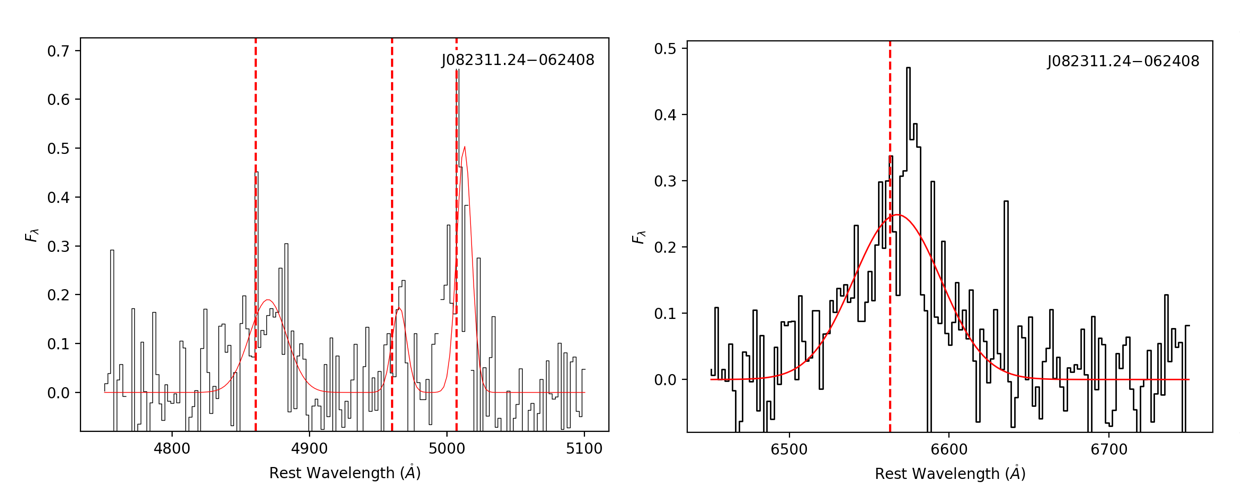}
\includegraphics[width = 0.9\textwidth]{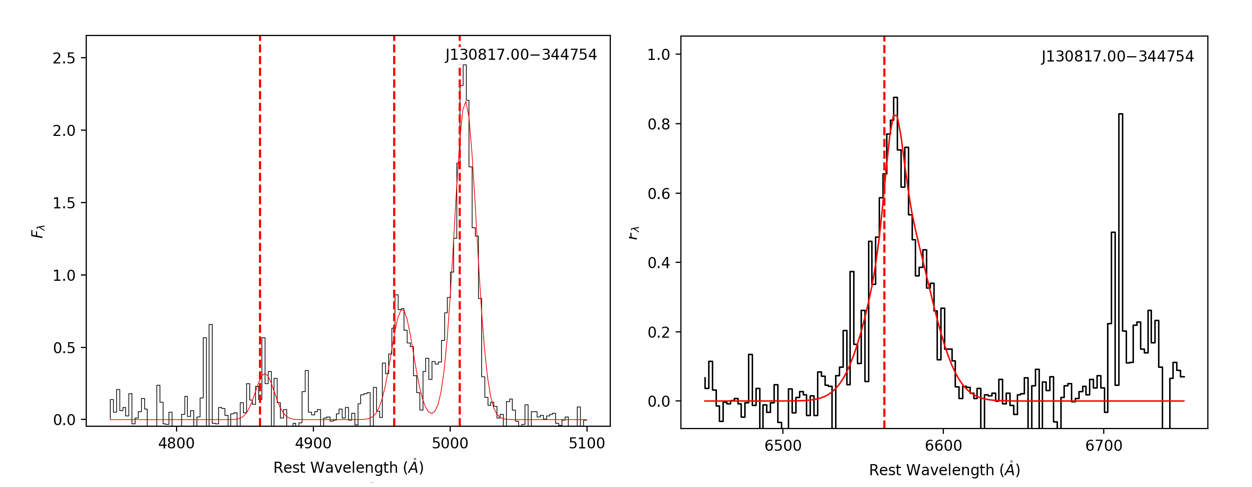}
\caption{\label{fig:specap} Rest frame X-shooter spectra in the region surrounding \hbeta{}, \oall{} and \halpha{}. Flux densities are in the units $10^{-16}$ $\rm{erg\,s^{-1}cm^{-2}\ang{}^{-1}}$. Reduced spectra and fitted model are represented by black histogram and red lines respectively. For \hbeta{} and \oall{} spectra the red line represents a combined model of the multi-Gaussian model fitted to each individual peak. For the \halpha{} spectra the red line also represents a multi-Gaussian model fitted to the blended \halpha{} and \nii{} profile. Vertical dashed red lines indicate the rest frame wavelengths of the \hbeta{}, \oall{} and \halpha{} emission lines based on the redshifts from \citet{Lonsdale2015}. Note that the redshift of W1702 was too high for the \halpha{} band to be observed.}
\end{figure*}

\begin{figure*}
\includegraphics[width = 0.9\textwidth]{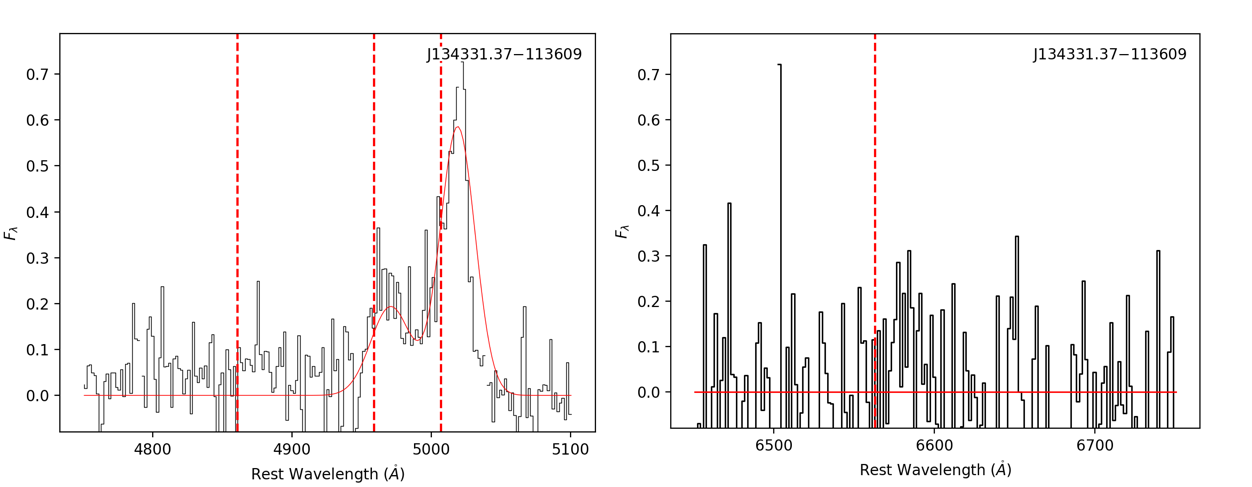}
\includegraphics[width = 0.9\textwidth]{figs/spec/1400.png}
\includegraphics[width = 0.9\textwidth]{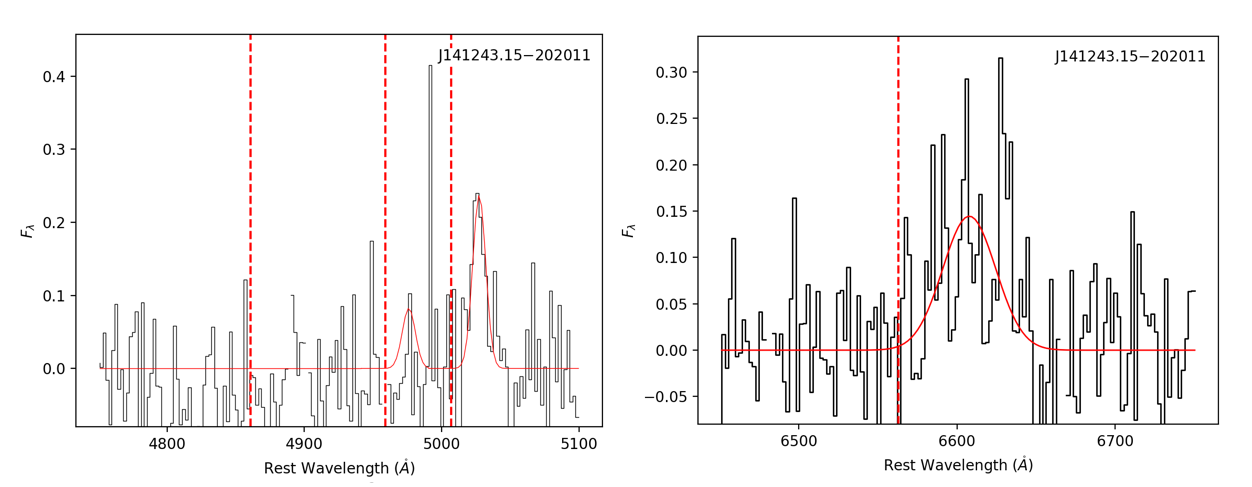}
\includegraphics[width = 0.9\textwidth]{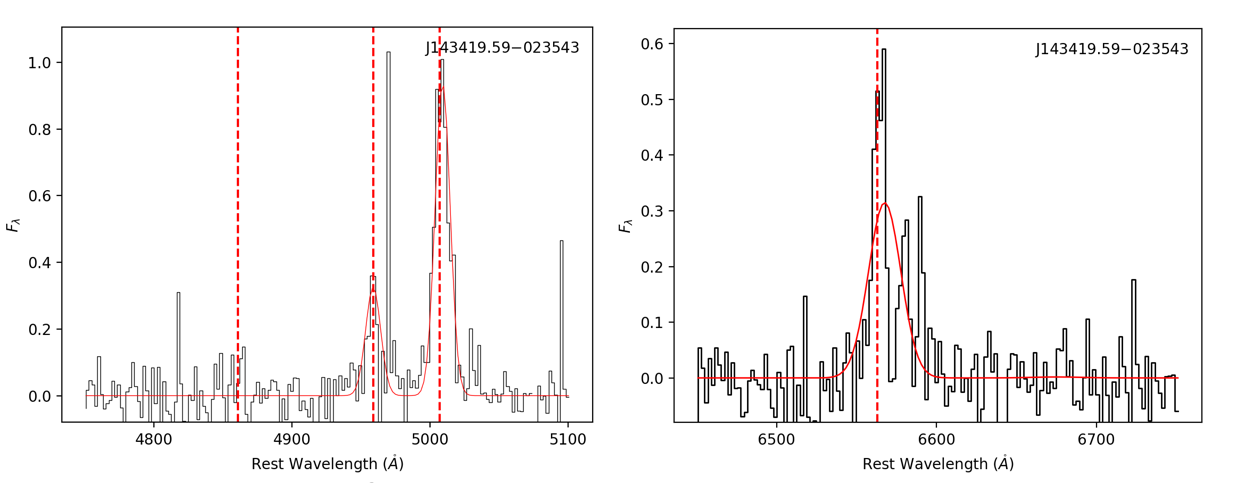}
\contcaption{}
\end{figure*}

\begin{figure*}
\includegraphics[width = 0.9\textwidth]{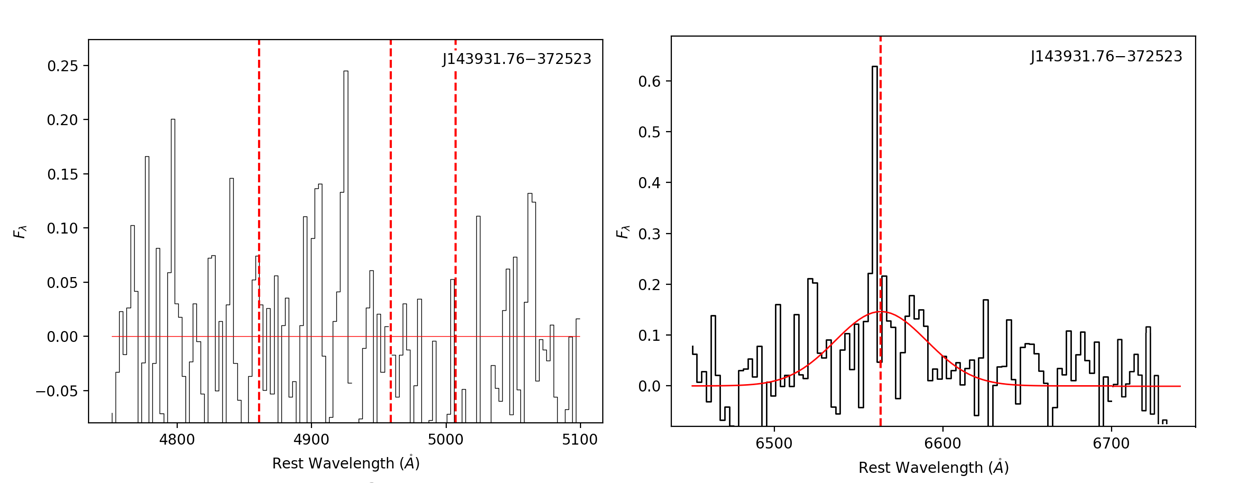}
\includegraphics[width = 0.9\textwidth]{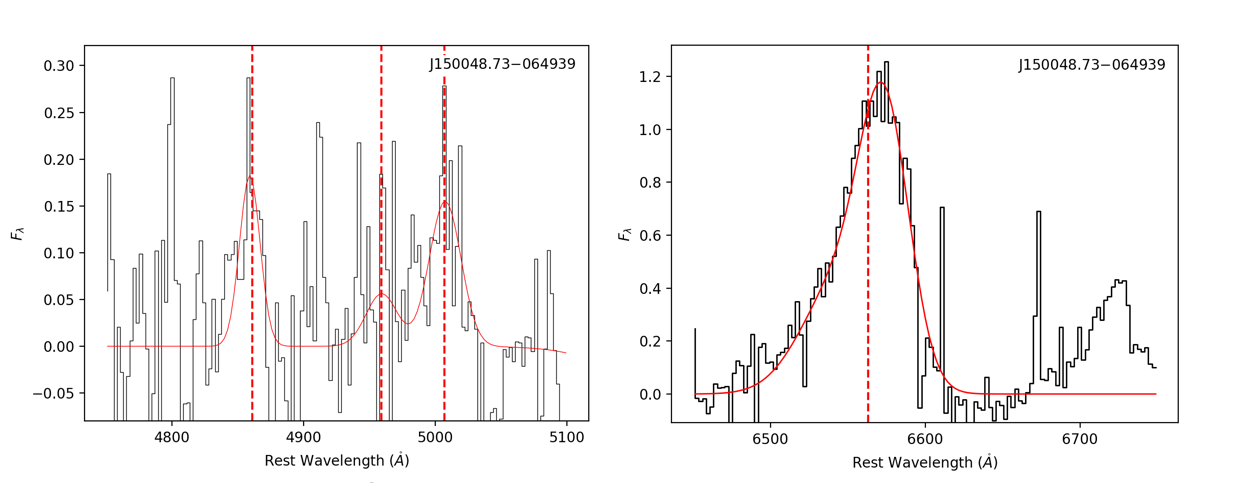}
\includegraphics[width = 0.9\textwidth]{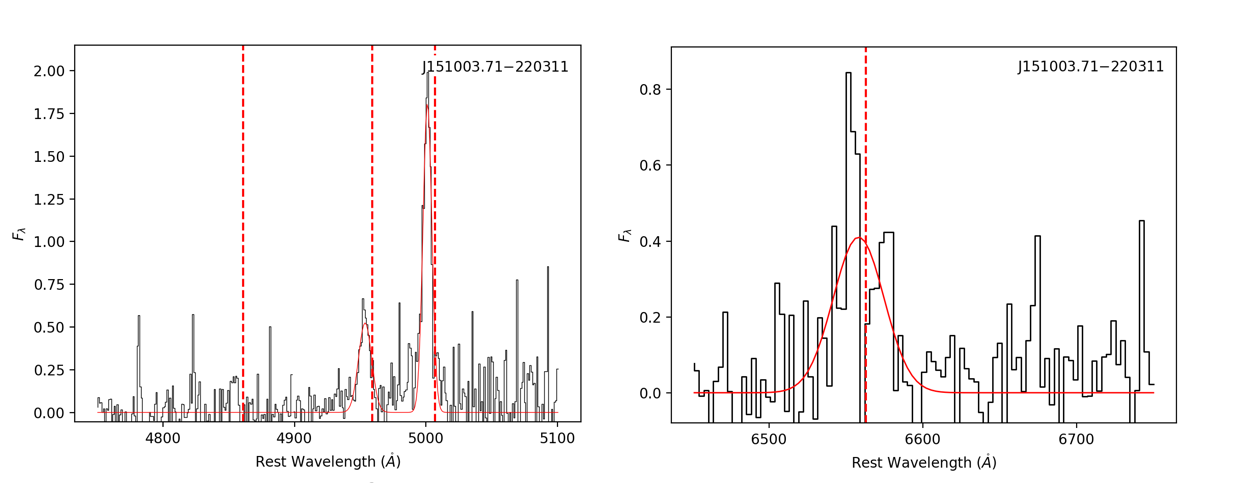}
\includegraphics[width = 0.9\textwidth]{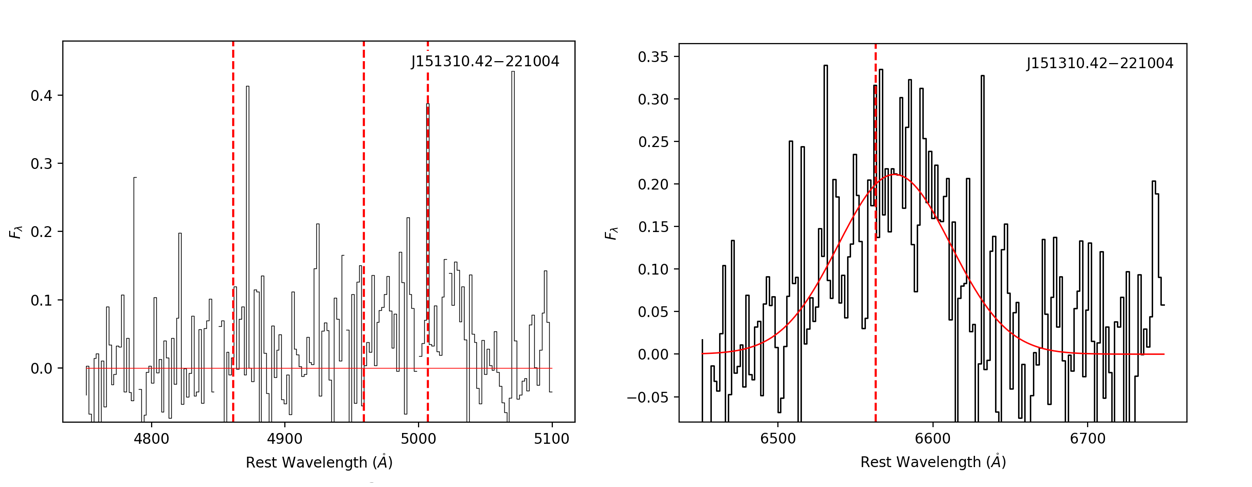}

\contcaption{}
\end{figure*}

\begin{figure*}
\includegraphics[width = 0.9\textwidth]{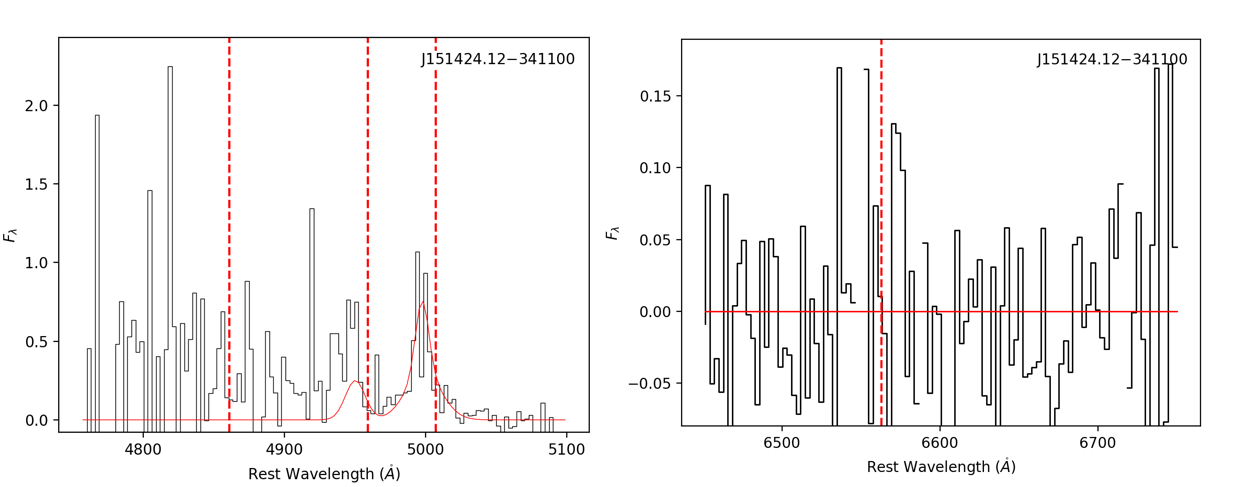}
\includegraphics[width = 0.9\textwidth]{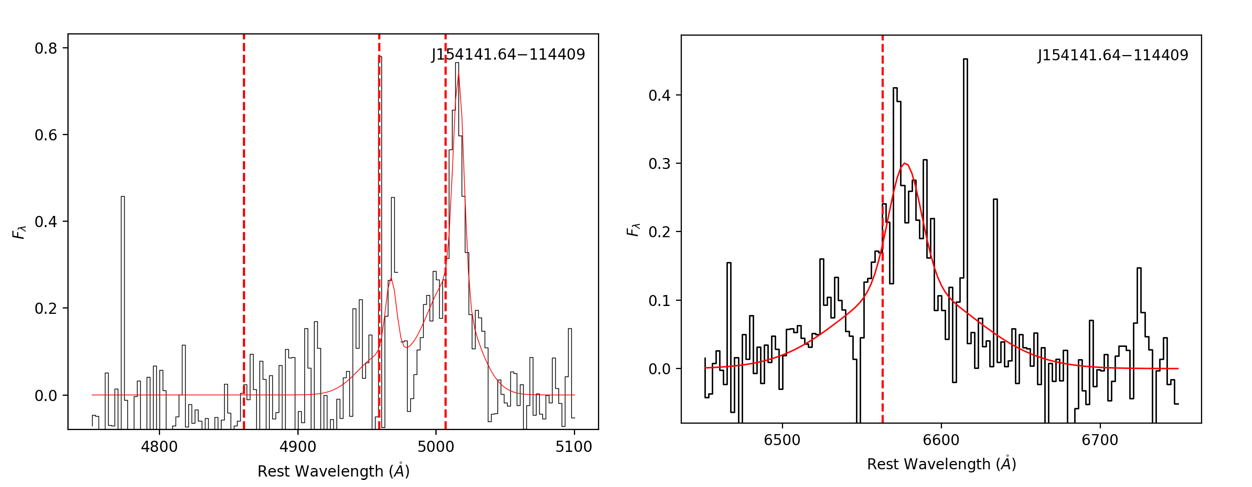}
\includegraphics[width = 0.9\textwidth]{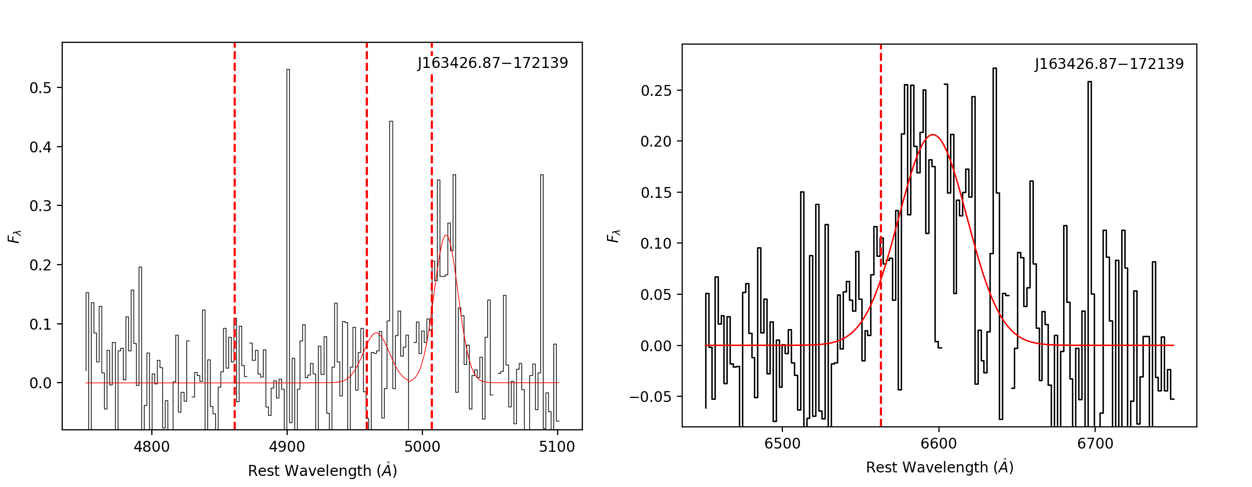}
\includegraphics[width = 0.9\textwidth]{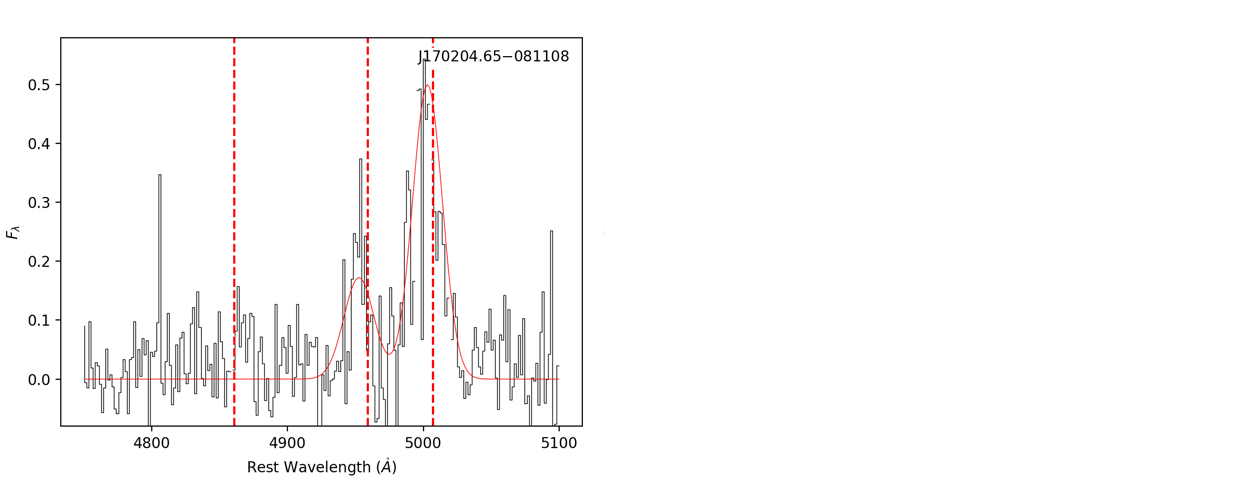}

\contcaption{}
\end{figure*}

\begin{figure*}
\includegraphics[width = 0.9\textwidth]{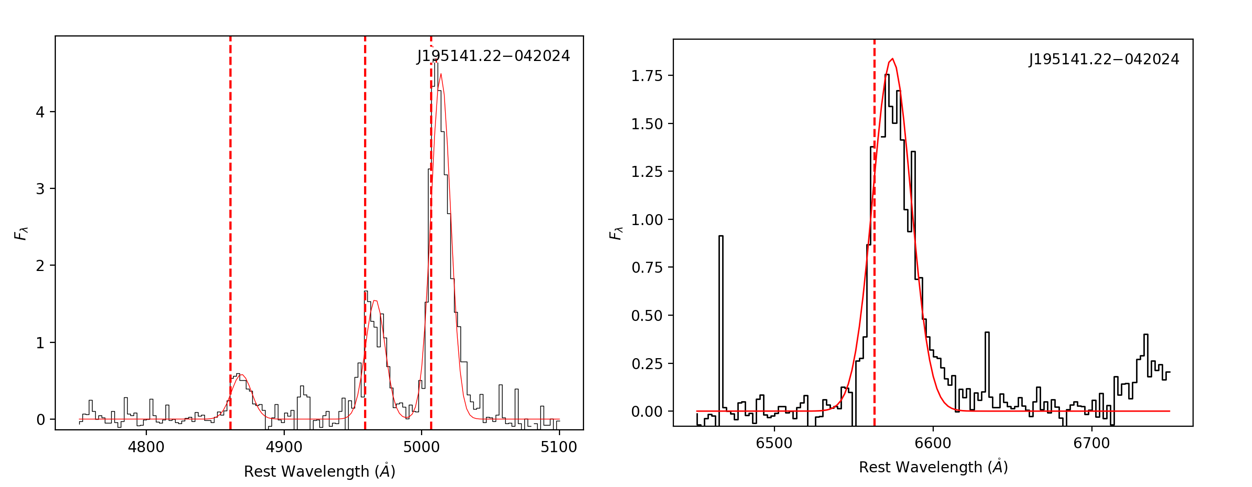}
\includegraphics[width = 0.9\textwidth]{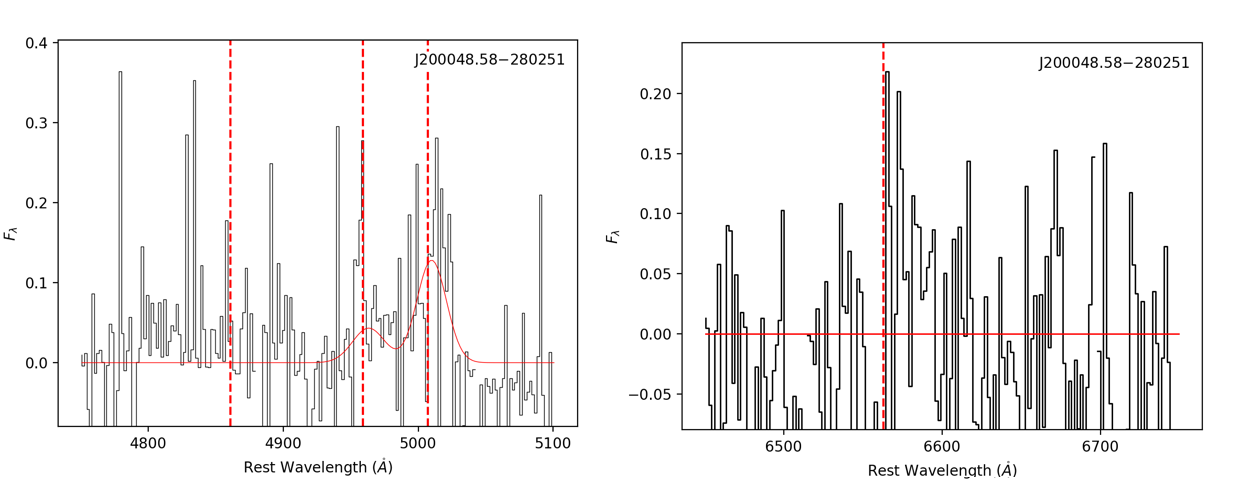}
\includegraphics[width = 0.9\textwidth]{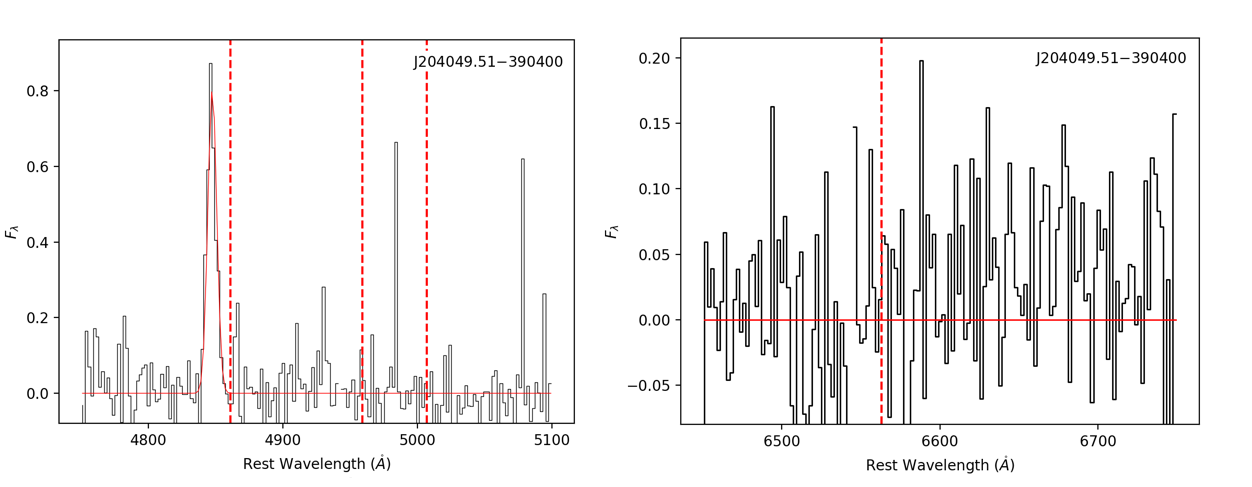}
\contcaption{}
\end{figure*}

\section{1D spectra with no line detections}
\label{sec:appbad}
Given below in Fig. \ref{fig:specapbad} are the full 1D spectra for the 9 sources with no detected emission lines. Note that bins with a flux error $\geq\,3 \times$ their value are masked.

\begin{figure*}
\centering
\includegraphics[width = 0.9\textwidth]{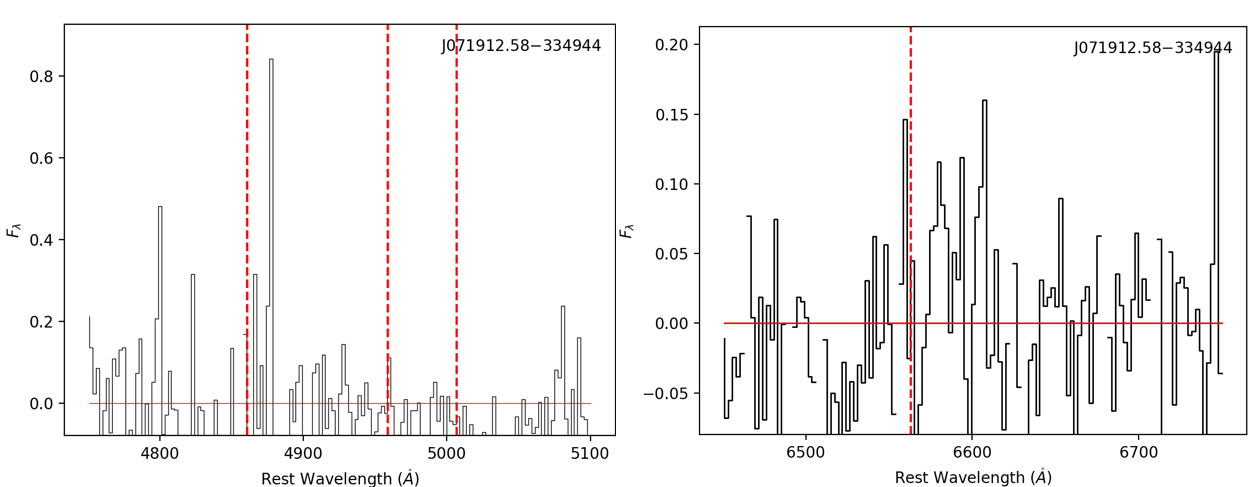}
\includegraphics[width = 0.9\textwidth]{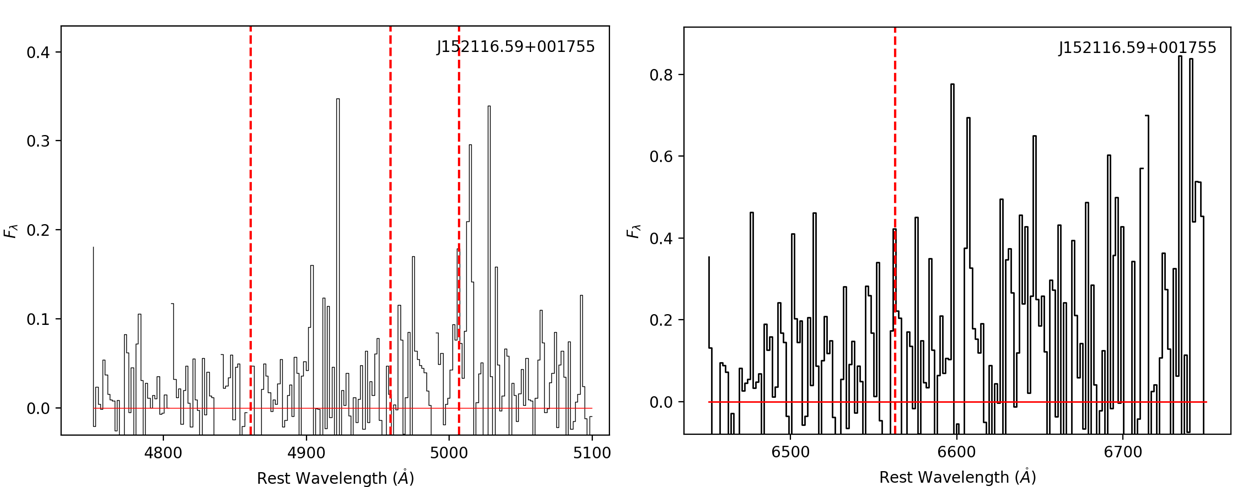}
\includegraphics[width = 0.9\textwidth]{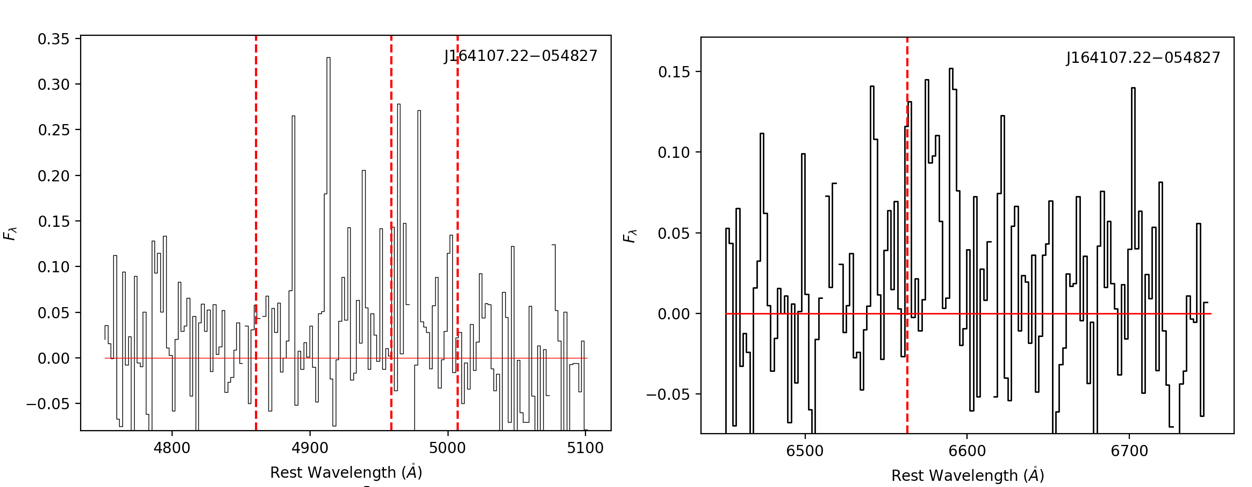}
\caption{\label{fig:specapbad} Rest frame X-shooter spectra in the region surrounding \hbeta{}, \oall{} and \halpha{} for the 9 sources with no detected emission lines. Flux densities are in the units $10^{-16}$ $\rm{erg\,s^{-1}cm^{-2}\ang{}^{-1}}$. Vertical dashed red lines indicate the rest frame wavelengths of the \hbeta{}, \oall{} and \halpha{} emission lines based on the redshifts from \citet{Lonsdale2015}. }
\end{figure*}

\begin{figure*}
\includegraphics[width = 0.9\textwidth]{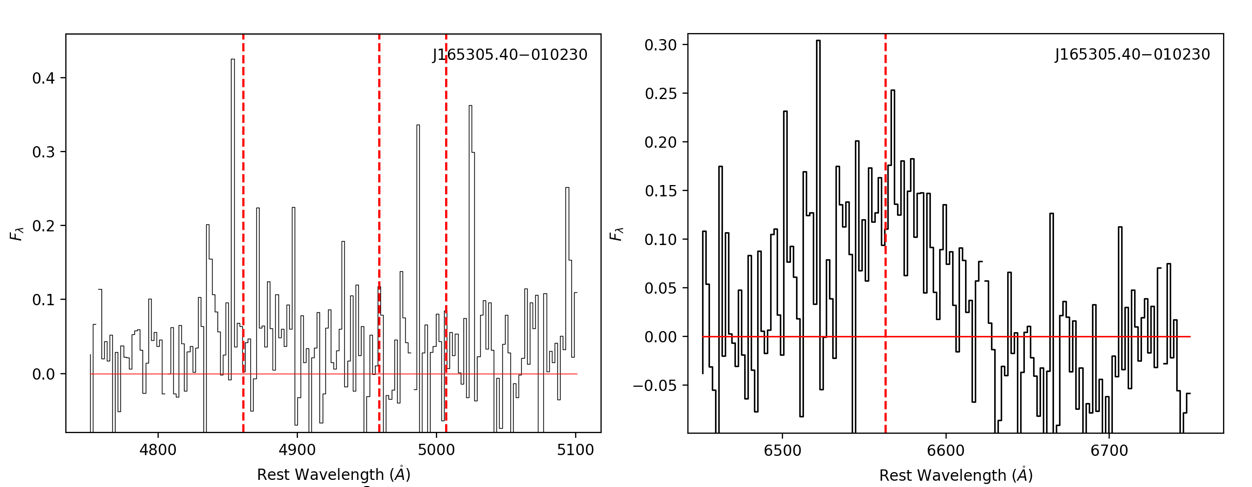}
\includegraphics[width = 0.9\textwidth]{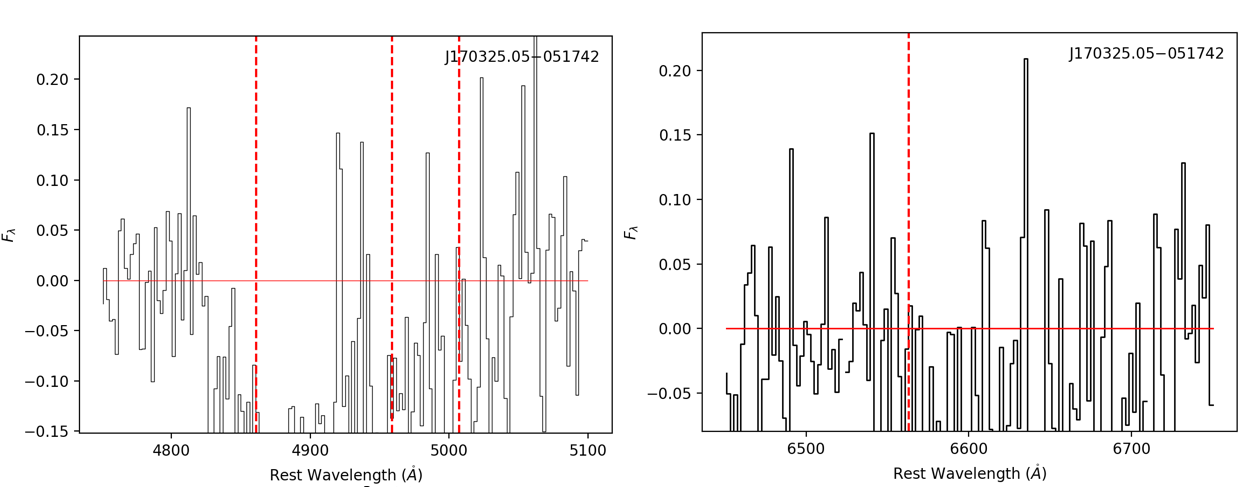}
\includegraphics[width = 0.9\textwidth]{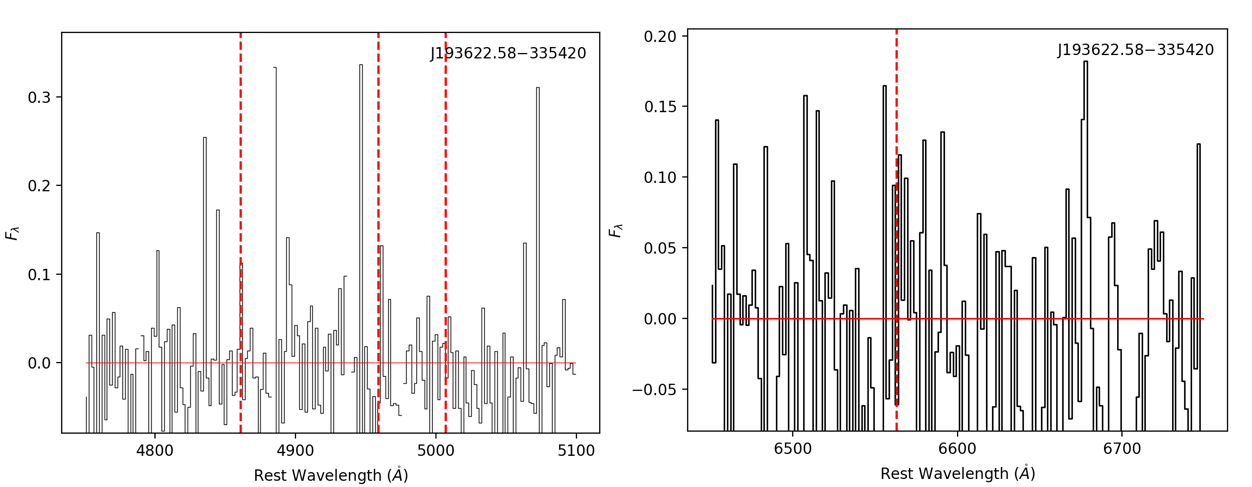}
\includegraphics[width = 0.9\textwidth]{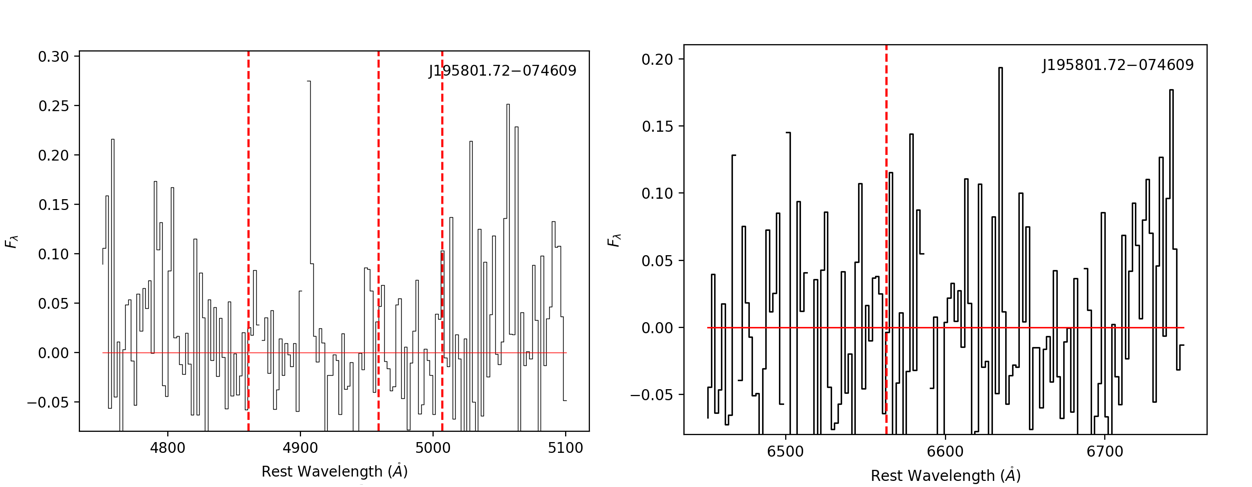}
\contcaption{}
\end{figure*}

\begin{figure*}
\includegraphics[width = 0.9\textwidth]{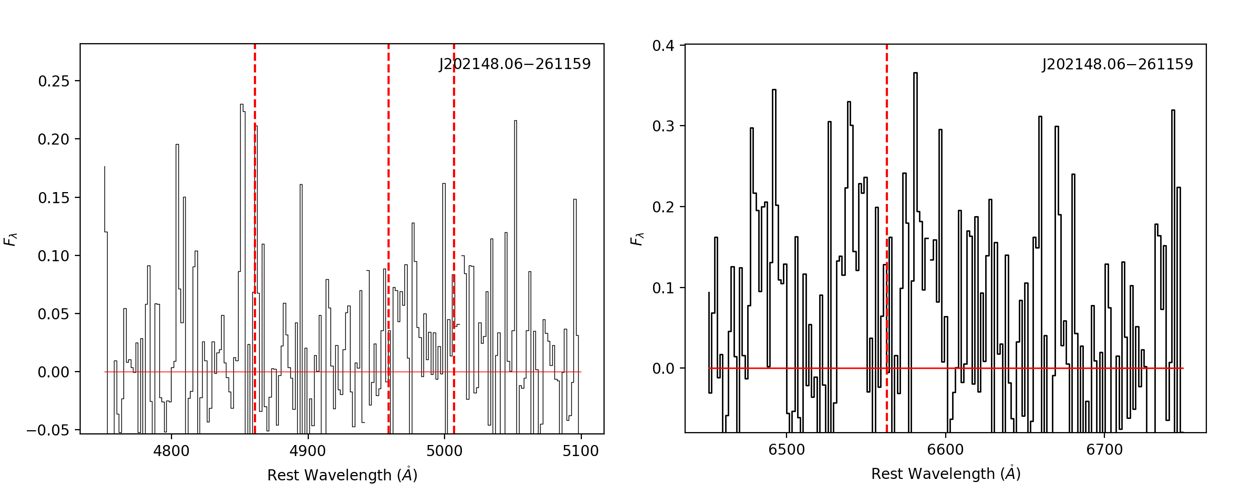}
\includegraphics[width = 0.9\textwidth]{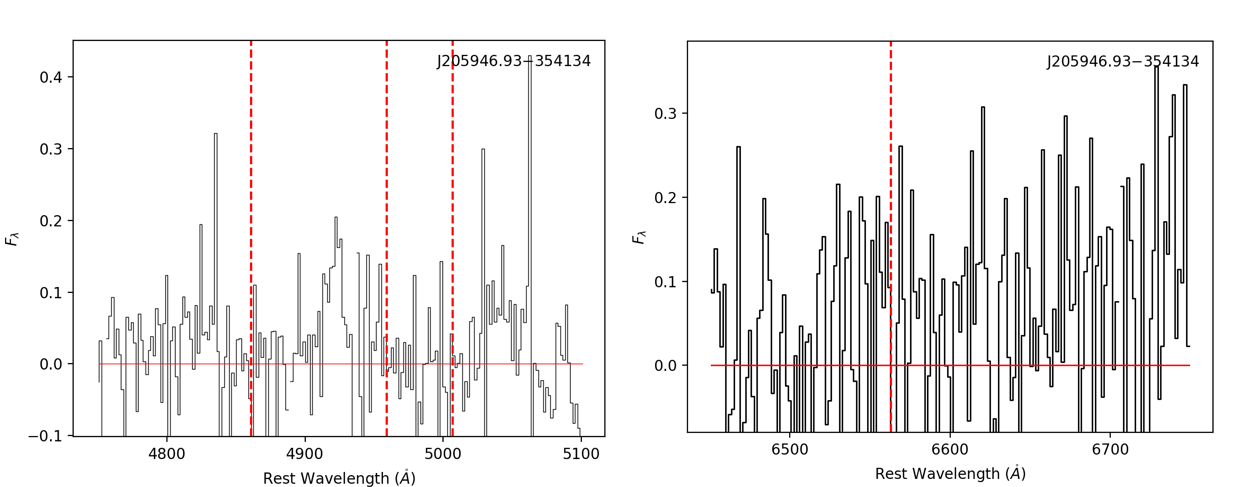}
\end{figure*}

\bsp	
\label{lastpage}
\end{document}